\documentclass[aps,prd,amsmath,amssymb,eqsecnum,nofootinbib,notitlepage]{revtex4-1}

\usepackage{hyperref}
\usepackage[retainorgcmds]{IEEEtrantools}
\usepackage{graphicx}
\usepackage{graphics}
\usepackage{color}

\allowdisplaybreaks[1]

\newcommand{\ab}{{\bar{a}}}
\newcommand{\bb}{{\bar{b}}}

\newcommand{\hb}{{\bar{h}}}

\newcommand{\xb}{{\bar{x}}}

\newcommand{\rb}{r_0}
\newcommand{\rbdot}{\dot{r}_0}

\newcommand{\alphab}{{\bar{a}}}
\newcommand{\betab}{{\bar{b}}}

\newcommand{\zrho}{{\rho}}
\newcommand{\zrhoz}{{\rho_0}}

\newcommand{\adv}{{(adv)}}
\newcommand{\ret}{{(ret)}}
\newcommand{\sing}{{(S)}}
\newcommand{\reg}{{(R)}}

\newcommand{\lpow}[1]{[#1]}
\newcommand{\lnpow}[1]{[\scalebox{0.85}{-}#1]}

\newcommand{\BL}{ Boyer-Lindquist }

\DeclareMathOperator{\sgn}{sgn}

\begin{document}
\title{High-order expansions of the Detweiler-Whiting singular field in Kerr spacetime}
\author{Anna Heffernan}
\email{anna.heffernan@ucd.ie}
\affiliation{School of Mathematical Sciences and Complex \& Adaptive Systems Laboratory, University College Dublin, Belfield, Dublin 4, Ireland}

\author{Adrian Ottewill}
\email{adrian.ottewill@ucd.ie}
\affiliation{School of Mathematical Sciences and Complex \& Adaptive Systems Laboratory, University College Dublin, Belfield, Dublin 4, Ireland}

\author{Barry Wardell}
\email{barry.wardell@gmail.com}
\affiliation{School of Mathematical Sciences and Complex \& Adaptive Systems Laboratory, University College Dublin, Belfield, Dublin 4, Ireland}

\begin{abstract}
In a previous paper, we computed expressions for the Detweiler-Whiting singular
field of point scalar, electromagnetic and gravitational charges following a
geodesic of the Schwarzschild spacetime. We now extend this to the case of
equatorial orbits in Kerr spacetime, using
coordinate and covariant approaches to compute expansions of the singular
field in scalar, electromagnetic and gravitational cases.  As an application, we
give the calculation of previously unknown mode-sum regularization parameters.
We also propose a new application of high-order approximations to the singular
field, showing how they may be used to compute $m$-mode regularization
parameters for use in the $m$-mode effective source approach to self-force
calculations.
\end{abstract}

\maketitle

\section{Introduction}

The two-body problem in general relativity is a long-standing, open problem going
back to work by Einstein himself.  With recent advances in gravitational wave
detector technology, this age-old problem has been given a new lease of life. 
Some of the key sources expected to be seen by both space- and ground-based
gravitational wave detectors are black hole binaries. Accurate models of black
hole binaries are required for their successful detection by gravitational wave
detectors.  This development is today motivating numerical, analytical
and experimental relativists to work together with the goal of producing models
of the inspiral and merger of black hole binary systems.

In modeling black hole binaries, it is widely accepted that for the scenario of
an extreme mass ratio inspiral (EMRI), the self-force approach is the model of
choice.  EMRIs are expected to be seen by space-based detectors such
as NGO/eLISA \cite{AmaroSeoane:2012km}.  Although NGO/eLISA has recently been
postponed, the gravitational wave research community is confident in its
inevitable flight.  In the meantime, recent research has shown the applicability of
self-force calculations to other black hole binary configurations \cite{LeTiec:2011dp,LeTiec:2011bk},
extending the application of self-force to ground-based detectors
such as LIGO and VIRGO.

Within the self-force approach, one perturbatively solves for the motion of a small
body in the background of a massive black hole.
Formal derivations of the equations of motion of a small body, moving in a
curved spacetime, have settled on the idea of a well-defined singular-regular
split of the retarded field generated by the body
\cite{Mino:1996nk,Quinn:1996am,Quinn:2000wa,Gralla:2008fg,Harte:2008xq,Harte:2009uy,Pound:2009sm,Gralla:2009md,Harte:2009yr}.
Several practical self-force computation strategies have developed from these
formal derivations, all of which are based on the now-justified assumption that
the use of a distributional source is acceptable at first perturbative order. These
strategies broadly fall into three categories:
the \emph{mode-sum} approach \cite{Barack:1999wf,Barack:2001gx},
the \emph{effective source} approach \cite{Barack:2007jh,Vega:2007mc} and
\emph{Green function} approaches \cite{Anderson:2005gb,Casals:2009zh}.
The key to all three approaches is the subtraction of an appropriate \emph{singular} component
from the retarded field to leave
a finite \emph{regular} field that is solely responsible for the self-force. This singular
component must have the same singular structure as the full retarded field in the vicinity
of the body and must not contribute to the self-force (or its contribution must be well known
such that it can be corrected for).
There are many choices for a singular field that satisfies these criteria,
although not all choices are equal. Detweiler and Whiting
\cite{Detweiler:2002mi} identified a particularly appropriate choice.
Through a Green function decomposition, they defined a singular field that not
only satisfies the above two criteria, but also has the property that when it is
subtracted from the full retarded field, it leaves a regularized field that is
a solution to the \emph{homogeneous} wave equation. Extensions of this idea of a
singular-regular split to extended charge distributions
\cite{Harte:2008xq,Harte:2009uy}, to second perturbative order
\cite{Rosenthal:2005ju,Rosenthal:2006iy,Detweiler:2011tt,Pound:2012nt,Gralla:2012db}
and to fully nonperturbative contexts \cite{Harte:2011ku} have recently been
developed.

In a previous paper \cite{Heffernan:2012su} (from now on referred to as Paper
I), we focused our calculations on the Schwarzschild spacetime representing a nonrotating
black hole.  Although this is a possible physical scenario, it is believed that a
more astrophysically realistic or probable situation would be that of a Kerr or rotating black
hole spacetime.  One of the primary goals of the self-force
community is, therefore, the successful calculation of the self-force in Kerr
spacetime, with particular emphasis on the gravitational case.  To this end, we
now adapt our previous work from Paper I to the Kerr spacetime.

In Paper I, we developed approaches to computing highly accurate approximations to the
Detweiler-Whiting singular field of point scalar and electromagnetic charges as
well as that of a point mass. This was achieved through high-order series
expansions in a parameter $\epsilon$, which acts as a measure of distance from
the particle's world line. We also derived explicit expressions for the
case of geodesic motion in Schwarzschild spacetime. In this paper, we extend this
analysis to the case of eccentric, equatorial orbits in the Kerr spacetime. We
find that all of the methods developed in Paper I may be applied to the Kerr
case with little modification. Nevertheless, the detailed expressions are
significantly more complicated in the Kerr case. Since our method is largely the
same as that used in Paper I, we direct the reader there for full details and
give here only the expressions that differ.

In Paper I, as applications of our high-order expansions of the singular field, we
derived expressions that may be used to improve the accuracy of both the mode-sum
and effective source approaches to computing the self-force.
Since the effective source approach requires that the
source be evaluated in an extended region around the world line, numerical
evaluation can be time consuming, in particular when using high-order
expansions such as the ones produced in both this paper and Paper I.  Existing
calculations have settled on expansions of the singular field to
$\mathcal{O}(\epsilon^2)$ as a particular
``sweet spot'' \cite{Dolan:2010mt,Dolan:2011dx,Diener:2011cc}
--- up to this order the increase in
complexity of the singular field and corresponding effective source is rewarded with an increase in
accuracy at modest computational cost.  However, expansions above this order may well slow down the calculations to
such a degree that the extra orders offer more of a hindrance than a help.
In this paper, we propose a solution to this problem that allows most of the benefit
to be reaped from high-order expansions without the need for using increasingly
complicated high-order expansions in numerical evolutions.
This idea makes use of the $m$-mode scheme, developed by Barack and Golbourn
\cite{Barack:2007jh},  which decomposes the retarded field and effective
source into azimuthal modes; the resulting conservation of axial symmetry makes the
scheme well suited
to the Kerr spacetime.  By carrying out $m$-mode effective source calculations with an
effective source accurate to some order, say $\mathcal{O}(\epsilon^2)$, one can obtain numerical values, for which the
$m$-modes of the self-force converge polynomially with $1/m$.  Our technique then makes use
of our higher terms of
the singular field [those above  $\mathcal{O}(\epsilon^2)$], to obtain a faster convergence of
this $m$-mode sum and, hence, assist in the production of highly accurate values
for the self-force.

The layout of this paper is as follows. In Sec.~\ref{sec:ModeSum}, we use
coordinate expansions to derive high-order regularization parameters for use in
the mode-sum method. In doing so we give new, previously unknown
regularization parameters in scalar, electromagnetic and gravitational
cases. In Sec.~\ref{sec:m-mode}, we propose a new application of high-order
coordinate expansions of the singular field, showing how they may be used
to derive $m$-mode regularization parameters for use in the $m$-mode
effective source approach. In Sec.~\ref{sec:Discussion}, we summarize our results
and discuss further prospects for their application.

Throughout this paper, we use units in which $G=c=1$ and adopt the sign conventions of
\cite{Misner:1974qy}. We denote symmetrization of indices using parenthesis
[e.g., $(ab)$], antisymmetrization using square brackets (e.g., $[ab]$), and exclude indices
from (anti)symmetrization by surrounding them by vertical
bars (e.g., $(a | b | c)$, $[a | b | c]$). We denote pairwise (anti)symmetrization using an
overbar, e.g., $R_{(\overline{ab}\,\overline{cd})} = \frac12 (R_{abcd}+R_{cdab})$. Partial derivatives
are represented by a comma (``,'') and covariant derivates by a semicolon (``;'').
Capital letters are used to denote the
spinorial/tensorial indices appropriate to the field being considered.
In many of our calculations, we have several spacetime points to be considered. Our convention
is that
\begin{itemize}
\item the point $x$ refers to the point where the field is evaluated,
\item the point $\xb$ refers to an arbitrary point on the worldline,
\item the point $x'$ refers to an arbitrary spacetime point,
\item the point $x_{\rm \adv}$ refers to the advanced point of $x$ on the world line,
\item the point $x_{\rm \ret}$ refers to the retarded point of $x$ on the world line.
\end{itemize}
In computing expansions, we use $\epsilon$ as an expansion parameter to denote the fundamental
scale of separation, so that $\Delta x = x-\xb \approx \mathcal{O}(\epsilon)$. Where
tensors are to be evaluated at these points, we decorate their indices appropriately using an overbar
$(\bar{~})$, e.g., $T^a$ and $T^\ab$ refer to tensors at $x$ and $\xb$,
respectively.

\section{\texorpdfstring{$\ell$}{l}-mode regularization}
\label{sec:ModeSum}
%: ----------------------- contents from here ------------------------

% ---------------------------------------------------------------------------
% -------------- Mode-Sum Decomposition of the Singular Field --------------------
% ---------------------------------------------------------------------------

One of the most successful self-force computation approaches to the date is the
\emph{mode-sum} scheme of Barack and Ori \cite{Barack:1999wf,Barack:2001gx}; the majority
of existing calculations are
based on it in one form or another
\cite{Barack:2000zq,Burko:2000xx,Detweiler:2002gi,DiazRivera:2004ik,Haas:2006ne,Haas:2007kz,Canizares:2009ay,Canizares:2010yx,Barack:2007tm,Barack:2002ku,Sago:2008id,Detweiler:2008ft,Sago:2009zz,Barack:2010tm,Keidl:2010pm,Shah:2010bi,Warburton:2010eq,Warburton:2011hp,Thornburg:2010tq,Haas:2011np,Warburton:2011fk,Hopper:2010uv}.
The basic idea is to decompose the retarded field into spherical harmonic modes,
which are continuous and finite - in general for the scalar case and in the Lorenz
gauge for the electromagnetic and gravitational cases. The spherical symmetry of
the Schwarzschild spacetime makes this decomposition into
spherical harmonic modes a natural choice. In Kerr spacetime, despite there
being more natural choices (such as a decomposition into spheroidal harmonics),
a decomposition of the singular field into spherical harmonic modes has been shown
to be of practical use in computing the scalar self-force
\cite{Warburton:2010eq,Warburton:2011hp}. While similar approaches have yet
to be attempted in electromagnetic or gravitational cases, it seems likely that they
are at least possible in principle.

A key component of the
mode-sum calculation involves the subtraction of the so-called \emph{regularization
parameters} -- analytically derived expressions that render the formally
divergent sum over spherical harmonic modes finite.  In this section, we derive
these parameters from our singular field expressions and show how they may be
used to compute the self-force with unprecedented accuracy.

% -------------------------------------------------------------------------------
% ------------------------------ Mode-Sum Concept -------------------------------
% -------------------------------------------------------------------------------

\subsection{Mode-sum concept}
The self-force for the scalar, electromagnetic and gravitational cases,
can be written generically as
\begin{equation} \label{eqn: Fa}
F^a = p^a{}_A \varphi^{A}_{\rm{\reg}},
\end{equation}
where
\begin{equation}
\varphi^{A}_{\rm{\reg}} = \varphi^{A}_{\rm{\ret}} - \varphi^{A}_{\rm{\sing}}
\end{equation}
is the regularized field and $p^a{}_A(x)$ is a tensor at $x$, which depends on the
type of charge.  We can, therefore, rewrite the self-force as
\begin{equation}
F^a = p^a{}_A \varphi^{A}_{\rm{\ret}} - p^a{}_A \varphi^{A}_{\rm{\sing}}.
\end{equation}
Carrying out a spherical harmonic decomposition on the field,
\begin{equation}
\varphi^{A}_{\rm{\ret / \sing}} = \sum_{\ell=0}^{\infty} \sum_{m=-\ell}^{\ell} \varphi_{\ell m}^{A}{}_{\rm{\ret / \sing}} Y_{\ell m} (\theta, \phi),
\end{equation}
allows the self-force to be rewritten as,
\begin{equation}
F^a = \sum_{\ell=0}^{\infty} \sum_{m=-\ell }^{\ell } \left(p^a{}_A \varphi_{\ell m}^{A}{}_{\rm{\ret}} - p^a{}_A \varphi_{\ell m}^{A}{}_{\rm{\sing}} \right) Y_{\ell m} (\theta_0, \phi_0).
\end{equation}
Defining the $\ell$ component of the retarded or singular self-force to be
\begin{equation} \label{eqn: FretSing}
F^a_{\ell}{}_{\rm \ret / \sing} = p^a{}_A  \sum _{m=-\ell }^{\ell } \varphi_{\ell m}^{A}{}_{\rm{\ret / \sing}} Y_{\ell m} (\theta_0, \phi_0),
\end{equation}
the self-force can be expressed as
\begin{equation} \label{eqn: FaSplit}
F^a = \sum_{\ell=0}^{\infty} \left( F^a_{\ell}{}_{\rm \ret}  -  F^a_{\ell}{}_{\rm \sing}  \right).
\end{equation}
It is the last term on the right, $F^a_\ell{}_{\rm \sing}$, that we calculate in this section for each of
the scalar, electromagnetic and gravitational cases in Kerr spacetime.

Our explicit expression for the $\ell$-modes of the singular self-force in Kerr
spacetime is written as an expansion about the world-line point $\bar{x}$, that
is
\begin{equation}
 F^a_{\ell}{}_{\rm \sing} = F^{\ell}_{a\lnpow{1}}\left(r_0,t_0\right) + F^{\ell}_{a[0]}\left(r_0,t_0\right) + F^{\ell}_{a\lpow{2}}\left(r_0,t_0\right)
	+\: F^{\ell}_{a\lpow{4}}\left(r_0,t_0\right) + F^{\ell}_{a\lpow{6}}\left(r_0,t_0\right) +  \dots,
\end{equation}
where we are missing odd orders above $-1$, as these are zero -
this will be shown to be the case later in this section.  When summed over $\ell$,
the contribution of $ F^a_{\ell \lpow{2}}\left(r_0,t_0\right)$ and higher terms to
the self-force is zero.  However, if we ignore these higher terms in the
approximation of $\varphi_{\ell m}^{A}{}_{\rm{\sing}}$, then the approximation for
$\varphi_{\ell m}^{A}{}_{\rm{\reg}}$ is only $C^1$, causing the sum over $\ell$ to be
polynomially, rather than exponentially convergent in $1/\ell$. Therefore, despite
these terms having zero total contribution to the self-force, when it comes to
numerically calculating the self-force using a finite number of $\ell$-modes, the
inclusion of the higher order terms dramatically reduces the number of modes
required and, hence, computation time. For this reason, every extra term or
regularization parameter that can be calculated is important.

% -------------------------------------------------------------------------------
% ------------------------------ Rotated Coordinates -------------------------------
% -------------------------------------------------------------------------------

\subsection{Rotated coordinates}
To obtain expressions that are readily written as mode-sums, previous
calculations
\cite{Barack:1999wf,Detweiler:2002gi,Haas:2006ne}
found it useful to work in a rotated coordinate frame.  In Paper I, we found it
most efficient to carry out this rotation prior to doing any calculations;
this also holds in the Kerr case.  To this end, we introduce
coordinates on the 2-sphere at $\xb$ in the form
\begin{equation}
w_{1} = 2 \sin\left(\frac{\alpha}{2}\right) \cos\beta, \quad\quad w_2 = 2 \sin\left(\frac{\alpha}{2}\right) \sin\beta,
\end{equation}
where $\alpha$ and $\beta$ are rotated angular coordinates given by
\begin{eqnarray}
\sin \theta \cos \phi &=& -\cos\alpha \sin \theta_0 - \sin\alpha \sin\beta \cos \theta_0, \\
\sin \theta \sin \phi &=& \sin \alpha \cos \beta, \\
\cos \theta &=& \cos\alpha \cos \theta_0 - \sin\alpha \sin\beta \sin \theta_0.
\end{eqnarray} 
The Kerr metric in these coordinates (with $\bar{x}$ chosen to
lie on the equator, i.e., at $\alpha=0, \theta_0=\pi /2$) is given by the line element
\begin{align}
ds^2 = & \left[\frac{8 M r}{ 4 r^2 +a^2 w_2^2 \left(4 - w_1^2 - w_2^2 \right)} - 1 \right] dt^2
	+  \left[\frac{4 r^2 + a^2 w_2^2 \left(4 - w_1^2 - w_2^2 \right)}{4 \left(r^2 - 2 M r + a^2
	\right)} \right] dr^2 \nonumber \\
&
	-\: dt dw_1 \frac{4 a M r \left[8 - w_2^2 \left(6 - w_1^2 - w_2^2 \right) \right]}
	{\sqrt{4 - w_1^2 - w_2^2} \left[4 r^2 +a^2 w_2^2 \left(4 - w_1^2 - w_2^2 \right) \right]} 
	- dt dw_2 \frac{4 a M r w_1 w_2 \left(6 - w_1^2 - w_2^2 \right)}{\sqrt{4 - w_1^2 
	- w_2^2} \left[4 r^2 +a^2 w_2^2 \left(4 - w_1^2 - w_2^2 \right) \right]} 
	\nonumber \\
&	
	+\: \frac{1}{4 \left(4 - w_1^2 - w_2 ^2 \right) \left[4 - w_2^2 \left(4 - w_1^2 - 
	w_2^2 \right) \right]} \left[g_{w_1 w_1} dw_1^2 + 2 g_{w_1 w_2} dw_1 dw_2 + g_{w_2 w_2} dw_2^2\right],
\end{align}
where
\begin{IEEEeqnarray}{rCl}
g_{w_1 w_1} &=& w_1^2 w_2^2 \left[4 r^2 +a^2 w_2^2 \left(4 - w_1^2 - w_2^2 
	\right) \right] + \left[8 - w_2^2 \left(6 - w_1^2 - w_2^2 \right) \right]^2 \Bigg[r^2 + a^2 + 2 M a^2 r 
	\frac{4 - w_2^2 \left(4 - w_1^2 - w_2^2 \right) }{4 r^2 + a^2 w_2^2 \left(4 - w_1^2 
	- w_2^2 \right) } \Bigg], \nonumber \\
g_{w_1 w_2} &=& w_1 w_2 \left(w_1^2 + 2 w_2^2 - 4 \right) \big[4 r^2 +a^2 
	w_2^2 \big(4 - w_1^2 - w_2^2 \big) \big] \nonumber \\
&&
	+\: w_1 w_2 \left(6 - w_1^2 - w_2^2 \right) \left[8 - w_2^2 
	\left(6 - w_1^2 - w_2^2 \right) \right] \Bigg[r^2 + a^2+ 2 M a^2 r \frac{4 - w_2^2 \left(4 - w_1^2 - w_2^2 \right) }{4 r^2 + a^2 w_2^2 
	\left(4 - w_1^2 - w_2^2 \right) } \Bigg], \nonumber \\
g_{w_2 w_2} & = & \left(4 - w_1^2 - w_2^2 \right)^2 \big[4 r^2 +a^2 
	w_2^2 \big(4 - w_1^2 - w_2^2 \big) \big] \nonumber \\
&&
	+\: w_1^2 w_2^2 \left(6 - w_1^2 - w_2^2 \right)^2 \Bigg[r^2 + a^2 
	+ 2 M a^2 r \frac{4 - w_2^2 \left(4 - w_1^2 - w_2^2 \right) }{4 r^2 + a^2 w_2^2 
	\left(4 - w_1^2 - w_2^2 \right) } \Bigg].
\end{IEEEeqnarray}
As in the Schwarzschild case, this algebraic form has an advantage over its
trigonometric counterpart in computer algebra programs where trigonometric
functions tend to slow down calculations.  Despite the apparent complexity of
the Kerr metric in this form, calculations of the regularization parameters using this form are
more efficient than using Boyer-Lindquist coordinates and rotating the resulting
complicated expressions.

% -------------------------------------------------------------------------------
% ------------------------------- Mode Decomposition -----------------------------
% -------------------------------------------------------------------------------

\subsection{Mode decomposition}
Having calculated the singular field using the Kerr metric in the above form and
the methods described in Paper I, it is straightforward to calculate the
singular component of the self-force,
$F^a$, for the scalar, electromagnetic and gravitational cases.  This is done by using
Eq.~\eqref{eqn: Fa} with the singular field substituted for the
regular field\footnote{In this section, for notational convenience
we drop the implied $\sing$ superscript denoting ``singular'' as we are always referring
to the singular component.}.  We, then, obtain a multipole decomposition of $F^a$ by writing
\begin{equation}\label{eqn:falpha}
F^a \left(r, t, \alpha, \beta \right) = \sum_{\ell=0}^{\infty} \sum_{m=-\ell}^{\ell} F^a_{\ell m} \left(r, t \right) Y_{\ell m} \left( \alpha, \beta \right),
\end{equation}
where $Y^{\ell m} \left( \theta, \phi \right)$ are scalar spherical harmonics, and
accordingly
\begin{equation} \label{eqn:flm}
F^a_{\ell m} \left(r, t\right) = \int F^a \left(r, t, \alpha, \beta \right) Y^*_{\ell m} \left( \alpha, \beta \right) d \Omega.
\end{equation}
To calculate the $\ell$-mode contribution at $\xb = \left( t_0, r_0, \alpha_0,
\beta_0 \right)$, we have
\begin{equation} \label{eqn:falphal}
F^a_{\ell} \left(r_0, t_0 \right) = \lim_{\Delta r \rightarrow 0} \sum_{m=-\ell}^{\ell} F^a_{\ell m} \left(r_0+\Delta r, t_0 \right) Y_{\ell m} \left( \alpha_0, \beta_0 \right).
\end{equation}
With the particle on the pole in the rotated coordinate system,
$Y_{\ell m} \left( \alpha_0 = 0, \beta_0 \right) = 0$  for all $m \neq 0$.  This
also allows us, without loss of generality, to take $\beta_0 = 0$.  Taking
$\alpha_0$, $\beta_0$ and $m$ all to be equal to zero in Eq.~(\ref{eqn:falphal})
gives
\begin{IEEEeqnarray}{rCl} \label{eqn:fla}
F^a_{\ell} \left(r_0, t_0\right) &=& \lim_{\Delta r \rightarrow 0} \sqrt{\frac{2 \ell + 1}{4 \pi}} F^a_{\ell0} \left(r_0+\Delta r, t_0 \right) \nonumber \\
&=& 
	\frac{2 \ell + 1}{4 \pi} \lim_{\Delta r \rightarrow 0} \int F^a \left( r_0+\Delta r, t_0, \alpha, 
	\beta \right) P_{\ell} \left( \cos \alpha \right) d \Omega .
\end{IEEEeqnarray}

Using the methods of Paper I, a coordinate expansion of the singular self-force
$F^a \left( r, t, \alpha, \beta \right)$ may be written in the form
\begin{equation}
\label{eqn:fasum}
F^a \left( r, t, \alpha, \beta \right) = \sum_{n=1}^{\infty} \frac{B^{a(3 n -2)}}{ \zrho^{2 n + 1}} \epsilon^{n-3},
\end{equation}
where $B^{a(k)} = b^a_{a_1 a_2 \cdots a_k}(\bar{x}) \Delta x^{a_1} \Delta x^{a_2} \cdots \Delta x^{a_k}$ and $\zrho = \sqrt{(g_{\alphab \betab} u^{\alphab} \Delta x^b)^2 +g_{\alphab \betab} \Delta x^a \Delta x^b}$. In using Eq.~(\ref{eqn:fasum}) to determine the regularization
parameters, we only need to take the term in the sum at the appropriate order: $n=1$ for
$F^a_{\ell[-1]}$, $n=2$ for $F^a_{\ell[0]}$, etc.  Explicitly, in our rotated coordinates 
\begin{align}
\zrho \left(r, t, \alpha, \beta \right)^2 =&\frac{\Delta r^2 \rb \left[\rb \left(a^2 E^2-L^2 \right) 
	+ 2 M (L-a E)^2+E^2 \rb^3\right]}{\left(a^2-2 M \rb +\rb^2\right)^2}+\Delta t \Bigg[ 
	\Delta w_1 \Big(-\frac{4 a M}{\rb}- 2 E L\Big) -\frac{2 \Delta r E \rb^2 \rbdot}{a^2-2 M \rb+\rb^2}\Bigg]\nonumber \\
&
	+\Delta w_1^2 
	\left( \frac{2 a^2 M}{\rb}+a^2+L^2+\rb^2\right)+ \frac{2 \Delta r \Delta w_1 L \rb^2 \rbdot}{a^2 - 2 M \rb+\rb^2}+\Delta t^2 
	\left(E^2+\frac{2 M}{\rb}-1\right)+\Delta w_2^2 \rb^2,
\end{align}
where the $\alpha$, $\beta$ dependence is contained exclusively in $\Delta w_1$
and $\Delta w_2$.  Here, $E = -u_t$ and $L = u_{\phi}$ are the energy per unit mass
and angular momentum along the axis of symmetry respectively.  In particular,
taking $t = t_0$ ($\Delta t = 0$) allows us to write
\begin{IEEEeqnarray}{rCl}\label{eqn: rhotKerr}
\zrho \left(r, t_0, \alpha, \beta \right)^2 &=& \frac{\Delta r^2 \rb \left[E \rb \left( a^2 + \rb^2 
	\right) +2 a M (a E-L)\right]^2}{\left(a^2-2 M \rb+\rb^2\right)^2 \left[\rb \left( a^2 + L^2 
	\right) +2 a^2 M+\rb^3\right]} + \Delta w_2^2 \rb^2\nonumber \\
&&
	+\:\bigg(\frac{2 a^2 M}{\rb}+a^2+L^2+ \rb^2\bigg) \left[\Delta w_1 + \frac{ \Delta r L \rb^3 \rbdot}{\left(a^2-2 M \rb +\rb^2 
	\right) \left( 2 a^2 M+a^2 \rb+L^2 \rb + \rb^3 \right)} \right]^2.
\end{IEEEeqnarray}
For the mode-sum decomposition, it is favorable to work with
$\zrhoz\left(\alpha, \beta \right)^2 \equiv  \zrho\left(r_0, t_0, \alpha, \beta \right)^2$
in the form
\begin{equation} \label{eqn: rhoz}
\zrhoz\left(\alpha, \beta \right)^2 = 2 \left(1 - \cos{\alpha}\right) \zeta^2 \left(1 - k \sin^2{\beta} \right).
\end{equation}
This can be achieved by rewriting Eq.~\eqref{eqn: rhotKerr} with $\Delta r
\rightarrow 0$ as
\begin{equation}
\zrhoz\left(\alpha, \beta \right)^2  = \zeta^2 \Delta w_1 ^2+\rb^2\Delta w_2^2,
\end{equation}
where
\begin{equation}
\zeta^2 = L^2 + \rb^2 +\frac{2 a^2 M}{\rb} + a^2. 
\end{equation}
Rearranging gives
\begin{equation}
\zrhoz\left(\alpha, \beta \right)^2  = 2 \left(1 - \cos{\alpha}\right) \zeta^2 \left[1 - \left(\frac{\zeta^2 - \rb^2}{\zeta^2}\right) \sin^2{\beta} \right],
\end{equation}
which is equivalent to Eq.~\eqref{eqn: rhoz} with $k =
\frac{\zeta^2-\rb^2}{\zeta^2}$.  Defining $\chi(\beta) \equiv 1 - k \sin^2
\beta$, we can now rewrite our $\Delta w$'s in the alternate form
\begin{align}
\Delta w_1 ^2 &= 2 \left( 1-\cos{\alpha}\right) \cos^2 \beta = \frac{\zrhoz^2}{\zeta^2 \chi}\cos^2\beta= \frac{\zrhoz^2}{\left(\zeta^2-\rb^2\right)\chi}\left[k-(1-\chi)\right], \\
\Delta w_2^2 &= 2 \left( 1-\cos{\alpha}\right) \sin^2 \beta = \frac{\zrhoz^2}{\zeta^2 \chi}\sin^2\beta= \frac{\zrhoz^2}{\left(\zeta ^2 - \rb^2 \right)\chi} (1 - \chi).
\end{align}
It is worth noting that these expressions are equivalent to those in Paper I, but are written in a more general form here - we can recover the Paper I expressions (Schwarzschild spacetime) by setting $\zeta^2 = L^2 + \rb^2$.

Suppose, for the moment, that we may take the limit in Eq.~(\ref{eqn:fla})
through the integral sign, then, using our alternate forms, we have
\begin{equation}
\lim_{\Delta r \to 0} \frac{B^{a(3 n -2)}}{ \zrho^{2 n + 1}} \epsilon^{n-3} = \frac{b^a_{i_1 i_2 \dots i_{3n-2}}(\rb) \Delta w^{i_1} \Delta w^{i_2} \dots \Delta w^{i_{3n-2}}} { \zrhoz^{2 n + 1}} \epsilon^{n-3} = \zrhoz^{n -3 } \epsilon^{n-3} c^a_{(n)}(\rb,\chi).
\end{equation}
In \cite{Barack:1999wf}, it was shown that the integral and limit in
Eq.~\eqref{eqn:fla} are indeed interchangeable for all orders except the leading
order, $n=1$ term, where the limiting $\Delta r/\zrhoz^{3}$ would not be integrable.
Thus we find the singular self-force now has the form
\begin{align} \label{eqn:fa}
F^a_{\ell} \left(r_0, t_0 \right) =& \frac{2 \ell + 1}{4 \pi}\Bigg[\epsilon^{-2}  \lim_{\Delta r \rightarrow 0} \int\frac{B^{a(1)} \left(r, t_0, \alpha, \beta \right)}{\zrho^3 \left(r, t_0, \alpha, \beta \right)} P_{\ell} \left( \cos \alpha \right) d \Omega + \sum_{n=2}^{\infty} \epsilon^{n-3} \int \zrhoz^{n-3} c^a_{(n)} \left(r_0, \chi \right) P_{\ell} \left( \cos \alpha \right) d \Omega \Bigg]\nonumber\\
\equiv & F^a_{\ell\lnpow{1}}\left(r_0,t_0\right) + F^a_{\ell[0]}\left(r_0,t_0\right) + F^a_{\ell\lpow{2}}\left(r_0,t_0\right) + F^a_{\ell\lpow{4}}\left(r_0,t_0\right) + F^a_{\ell\lpow{6}}\left(r_0,t_0\right) +  \dots.
\end{align}
Here, the $\beta$ dependence in the $c^a_{(n)}$'s are hidden in $\chi$, while the
$\alpha$, $\beta$ dependence of $F^a \left(r, t_0, \alpha, \beta \right)$ is
hidden in both the $\zrho$'s and $c^a_{(n)}$'s. Note here that we use the convention
that a subscript in square brackets denotes the term that will contribute at
that order in $1/\ell$. Furthermore, the integrand in the summation is odd or even
under $\Delta w_i \to - \Delta w_i$ according to whether $n$ (and so $3n-2$) is odd or even. As a result only the even terms are nonvanishing, while
$F^a_{\ell\lpow{1}}\left(r_0,t_0\right) = F^a_{\ell\lpow{3}}\left(r_0,t_0\right)=
F^a_{\ell\lpow{5}}\left(r_0,t_0\right)=0$, etc.

Some care is required in order to obtain easily integrable expressions in the case
of eccentric orbits. We use the approach of previous methods
\cite{Detweiler:2002gi,Barack:1999wf,Mino:2001mq,Haas:2006ne}
(and also employed in Paper I), by redefining our $\Delta w_1$ coordinate
in such a way that the cross terms involving $\Delta r \Delta w_1$ in $\zrhoz$
vanish. That is, we make the replacement
$\Delta w_1 \rightarrow \Delta w_1 + c \Delta r$, where $c$ is given by
\begin{IEEEeqnarray}{rCl}
c &=& \frac{ - L \rb^3 \rbdot}{\left(a^2-2 M \rb +\rb^2 \right) \left( 2 a^2 M+a^2 
	\rb+L^2 \rb + \rb^3 \right)}.
\end{IEEEeqnarray}
This allows us to write
\begin{align} \label{eqn: rhoForm}
\zrho \left(r, t_0, \alpha, \beta \right)^2 &= \nu^2 \Delta r^2 + \zeta^2 \Delta w_1^2 + \rb^2 \Delta w_2^2 \nonumber \\
&=  \nu^2 \Delta r^2 + 2 \chi \zeta^2 \left(1 - \cos{\alpha}\right)
\end{align}
where $\nu$ is an expression involving $\rb$, $a$, $E$ and $L$. This can easily be
rearranged to give
\begin{IEEEeqnarray}{rCl}
\zrho \left(r, t_0, \alpha, \beta \right)^{-3} &=& \zeta^{-3} \left(2 \chi \right)^{-3/2} \left(\delta^2 + 
	1 - \cos{\alpha} \right)^{-3/2} \nonumber \\
&=&
	\zeta^{-3} \left(2\chi \right)^{-3/2} \sum_{\ell=0}^{\infty} \mathcal{A}_{\ell}^{{-3}/{2}} (\delta) P_{\ell} \left( \cos \alpha \right),
\end{IEEEeqnarray}
where 
\begin{equation}
\delta^2 = \frac{\nu^2 \Delta r^2}{2 \zeta^2 \chi}   \quad \quad \text{and} \quad \quad  \mathcal{A}_{\ell}^{-\frac{3}{2}} (\delta) = \frac{2 \ell + 1}{\delta}
\end{equation} 
Here, $\mathcal{A}_{\ell}^{-\frac{3}{2}} (\delta)$ is derived from the generating
function of the Legendre polynomials as shown in Eq.~(D12) of
\cite{Detweiler:2002gi}. We can now express
$\zrho \left(r, t_0, \alpha, \beta \right)^{-3} $ as
\begin{equation}
\zrho \left(r, t_0, \alpha, \beta \right)^{-3} =  \frac{1}{\zeta^2 \nu \chi \sqrt{\Delta r^2}} \sum_{\ell=0}^{\infty} \left( \ell + \tfrac{1}{2} \right) P_{\ell} \left( \cos \alpha \right).
\end{equation}
Bringing this result into our expression for
$F^a_{\ell\lnpow{1}}\left(r_0,t_0\right)$ from Eq.~\eqref{eqn:fa} and integrating
over $\alpha$ gives
\begin{IEEEeqnarray}{rCl} \label{eqn: Fa1}
F^a_{\ell\lnpow{1}}\left(r_0,t_0\right) &=& \frac{1}{2 \pi} \left(\ell + \tfrac{1}{2} \right) \lim_{\Delta 
	r \rightarrow 0} \frac{1}{\zeta^2 \nu \sqrt{\Delta r^2}} \int \frac{\tilde{B}^{a(1)}}{\chi} \sum_{\ell'=0}^{\infty} \frac{2\ell+1}{2} P_{\ell} \left(\cos \alpha \right)P_{\ell'} \left(\cos \alpha \right)d\Omega \nonumber \\
&=&
	\left(\ell + \tfrac{1}{2} \right) \lim_{\Delta r \rightarrow 0} \frac{\tilde{b}^a_{a_r} \Delta r}
	{\zeta^2 \nu \sqrt{\Delta r^2}} \frac{1}{2 \pi} \int_0^{2\pi} \chi^{-1} d\beta \nonumber \\
&=&
	\left(\ell + \tfrac{1}{2} \right) \frac{\tilde{b}^a_{a_r} \sgn \left(\Delta r\right)} {\zeta \nu \rb}.
\end{IEEEeqnarray}
Here, the first equality takes advantage of the orthogonal nature of the $P_{\ell}
\left( \cos \alpha \right)$, while the last equality comes from taking the limit as
$\Delta r \rightarrow  0$ and noting from Appendix C of
\cite{Detweiler:2002gi} that the integral is a
special case of the hypergeometric functions given by
\begin{equation}
\frac{1}{2 \pi} \int \chi^{-1} d\beta  = F \left( 1, \tfrac{1}{2};1;k \right) = \frac{1}{\sqrt{1-k}} = \frac{\zeta}{\rb}.
\end{equation}
$B^{a(1)}$ and $b^a_{a_r}$ now also carry a tilde to signify that they are not
the exact same $B^{a(1)}$ and $b^a_{a_r}$ from Eq.~\eqref{eqn:fasum}; the tilde
reflects the fact that they have also undergone the coordinate shift $\Delta w_1
\rightarrow \Delta w_1 + c \Delta r$.  Again, it should be noted that Eq.~\eqref{eqn: Fa1} holds for any spacetime for which $\zrho = \sqrt{(g_{\alphab \betab} u^{\alphab} \Delta x^b)^2 +g_{\alphab \betab} \Delta x^a \Delta x^b}$ can be written in the form of Eq.~\eqref{eqn: rhoForm}.

In the higher order terms of Eq.~(\ref{eqn:fa}),
we may immediately work with  $\zrhoz^2= 2\chi \zeta^2 (1-\cos\alpha)$ so,
\begin{IEEEeqnarray}{rCl}
\zrhoz \left(r_0, t_0, \alpha, \beta \right)^n &=& \zeta^n \left[2\chi \left( 1 - 
	\cos \alpha \right)\right]^{{n}/{2}} \nonumber \\
&=&
	\zeta^n \left(2\chi \right)^{n/2} \sum_{\ell=0}^{\infty} \mathcal{A}_{\ell}^{{n}/{2}} (0) P_{\ell} 
	\left( \cos \alpha \right),
\end{IEEEeqnarray}
where $\mathcal{A}_{\ell}^{-\frac{1}{2}} (0) = \sqrt{2}$, from the generating function
of the Legendre polynomials and, as given in Appendix~D of
\cite{Detweiler:2002gi}, for $(n+1)/2\in\mathbb{N}$,
\begin{align}
\mathcal{A}^{n/2}_{\ell} \left(0 \right) =& 
		 \frac{\mathcal{P}_{n/2} \left( 2 \ell + 1\right) }{\left(2 \ell - n\right)\left(2 \ell - n +2\right) \cdots \left(2 \ell + n\right) \left(2 \ell + n +2 \right)},
\end{align}
where
\begin{align}
\mathcal{P}_{n/2} = &\left(-1\right)^{(n+1)/2} 2^{1 + n/2} \left( n!! \right)^2. \nonumber \\
\end{align}
In this case, the angular integrals involve
\begin{eqnarray} \label{eqn:zeta minus n}
\frac{1}{2 \pi} \int  \frac{d \beta}{\chi(\beta)^{n/2}} &=& 
\left<\chi^{-{n/2}}(\beta) \right> 
={}_2 F_1 \left(\frac{n}{2}, \frac{1}{2}; 1; k \right),
\end{eqnarray}
where $(n+1)/2\in\mathbb{N}\cup \{0\}$. The resulting equations can then be
tidied up using the following special cases of hypergeometric functions:
\begin{IEEEeqnarray}{lClClCl}
\left<\chi^{-\frac{1}{2}}\right> &=& \mathcal{F}_{\frac{1}{2}}(k) &=& {}_2F_1 \left(\frac{1}{2}, \frac{1}{2};1;k \right) &=& \frac{2}{\pi} \mathcal{K}, \\
\left<\chi^{\frac{1}{2}}\right> &=& \mathcal{F}_{-\frac{1}{2}}(k) &=& {}_2F_1 \left(-\frac{1}{2}, \frac{1}{2};1;k \right) &=& \frac{2}{\pi} \mathcal{E} ,
\end{IEEEeqnarray}
where
\begin{equation}
\mathcal{K} \equiv \int_0^{\pi/2} (1 - k \sin^2 \beta)^{-1/2} d\beta, \quad
\mathcal{E} \equiv \int_0^{\pi/2} (1 - k \sin^2 \beta)^{1/2} d\beta
\end{equation}
are complete elliptic integrals of the first and second kinds respectively. All
other powers of $\chi$ can be integrated to give hypergeometric functions, which
can then be manipulated to be one of the above by the use of the recurrence
relation in Eq.~(15.2.10) of \cite{Abramowitz:Stegun},
\begin{equation}
\mathcal{F}_{p+1} (k) = \frac{p-1}{p \left(k - 1\right)} \mathcal{F}_{p-1}(k) + \frac{1 - 2p + \left(p - \frac{1}{2}\right) k}{p \left(k - 1\right)} \mathcal{F}_p(k).
\end{equation}

In the following sections, we give the results of applying this calculation to each
of the scalar, electromagnetic and gravitational cases in turn. In doing so, we omit
the explicit dependence on $\ell$, which in each case is
\begin{gather}
F^a_{\ell\lnpow{1}} = (2\ell+1) F^a_{\lnpow{1}}, \quad F^a_{\ell[0]} = F^a_{[0]}, \quad F^a_{\ell\lpow{2}} = 
	\frac{F^a_{\lpow{2}}}{(2\ell-1)(2\ell+3)}, \nonumber \\
F^a_{\ell\lpow{4}} = \frac{F^a_{\lpow{4}}}{(2\ell-3)(2\ell-1)(2\ell+3)(2\ell+5)}, \nonumber \\
F^a_{\ell\lpow{6}} = \frac{F^a_{\lpow{6}}}{(2\ell-5)(2\ell-3)(2\ell-1)(2\ell+3)(2\ell+5)(2\ell+7)}.
\end{gather}

% -------------------------------------------------------------------------------
% --------------------------------- Scalar Case ------------------------------------
% -------------------------------------------------------------------------------

\subsection{Scalar \texorpdfstring{$\ell$}{l}-mode regularization parameters} \label{sec:scalar-regularization}
In the scalar case, the singular part of the self-force is given by
\begin{equation}
F_a = \partial_a \Phi^{\rm \sing},
\end{equation}
where $\Phi^{\rm \sing}$ is the scalar singular field.  The scalar regularization parameters are then given by
\begin{gather}
F_{t\lnpow{1}} = \frac{\rb \rbdot \sgn(\Delta r)}{2 [\rb \left(a^2+L^2\right)+2
   a^2 M+\rb^3]}, \nonumber \\
F_{r\lnpow{1}} = -\frac{\sgn(\Delta r) \left[E \rb \left(a^2+\rb^2\right)+2
   a M (a E-L)\right]}{2\left[a^2-2 M \rb+\rb^2\right]
   \left[\rb \left(a^2+L^2\right)+2 a^2 M+\rb^3\right]}, \nonumber \\
F_{\theta\lnpow{1}} = 0, \quad
F_{\phi\lnpow{1}} = 0,
\end{gather}
\begin{equation}
F_{t[0]} = \frac{\rbdot \left[F^{\mathcal{E}}_{t[0]} \mathcal{E} + F^{\mathcal{K}}_{t[0]} \mathcal{K} \right]}{\pi  \rb^2 \left[\rb^2+L^2 + \frac{2 a^2 M}{\rb} + a^2 \right]^{3/2}\Big[2 a^2 M + a^2 \rb + L^2 \rb \Big]^2} ,
\end{equation}
where
\begin{IEEEeqnarray*}{rCl}
F^{\mathcal{E}}_{t[0]} &=& 4 a L M \left(4 a^4 M^2+2 a^4 M \rb+2 a^2 L^2 M \rb-a^2 M \rb^3-a^2
   \rb^4-L^2 \rb^4\right) \nonumber \\
&&
	+\: E \big(-12 a^6 M^3-16 a^6 M^2
   \rb-7 a^6 M \rb^2-a^6 \rb^3-4 a^4 L^2 M^2 \rb-6 a^4 L^2 M
   \rb^2 - 2 a^4 L^2 \rb^3-6 a^4 M^2 \rb^3-5 a^4 M \rb^4 \\
&& \quad
	-\: a^4 \rb^5+a^2 L^4 M \rb^2-a^2 L^4 \rb^3 - 5 a^2 L^2 M \rb^4-3
   a^2 L^2 \rb^5-2 L^4 \rb^5\big), \\
F^{\mathcal{K}}_{t[0]} &=& -2 a
   L M \big(2 a^4 M^2-a^4 M \rb-a^4 \rb^2-a^2 L^2 M \rb-2 a^2 L^2
   \rb^2-2 a^2 M \rb^3-2 a^2 \rb^4 - L^4 \rb^2-2 L^2
   \rb^4\big) \nonumber \\
&&
	+\: E \big(4 a^6 M^3+4 a^6 M^2 \rb+a^6 M \rb^2-2 a^4 L^2 M^2
   \rb-a^4 L^2 M \rb^2+2 a^4 M^2 \rb^3 +a^4 M \rb^4-2 a^2 L^4
   M \rb^2+a^2 L^2 M \rb^4 \\
&& \quad
	+\: a^2 L^2 \rb^5+L^4 \rb^5\big),
\end{IEEEeqnarray*}
\begin{equation}
F_{r[0]} = \frac{F^{\mathcal{E}}_{r[0]} \mathcal{E} + F^{\mathcal{K}}_{r[0]} \mathcal{K}}{\pi  \rb^3 \left( 2 a^2 M + a^2 \rb + L^2 \rb \right)^2 \left(\rb^2+L^2 + \frac{2 a^2 M}{\rb} + a^2 \right)^{3/2} \left(\rb^2-2 M \rb +a^2 \right)},
\end{equation}
where
\begin{IEEEeqnarray*}{rCl}
F^{\mathcal{E}}_{r[0]} &=& \big(-24 a^8 M^3 \rb-32 a^8 M^2 \rb^2-14 a^8 M \rb^3-2 a^8
   \rb^4+24 a^6 L^2 M^4+12 a^6 L^2 M^3 \rb - 30 a^6 L^2 M^2 \rb^2-27
   a^6 L^2 M \rb^3 \\
&& \quad
	-\: 6 a^6 L^2 \rb^4+48 a^6 M^4 \rb^2+40 a^6 M^3
   \rb^3 - 16 a^6 M^2 \rb^4-20 a^6 M \rb^5-4 a^6 \rb^6+8 a^4
   L^4 M^3 \rb-12 a^4 L^4 M \rb^3 \\
&& \quad
	-\: 6 a^4 L^4 \rb^4 + 36 a^4 L^2 M^3
   \rb^3+12 a^4 L^2 M^2 \rb^4-21 a^4 L^2 M \rb^5-9 a^4 L^2
   \rb^6+24 a^4 M^3 \rb^5 + 8 a^4 M^2 \rb^6-6 a^4 M \rb^7 \\
&& \quad
	-\: 2 a^4 \rb^8-2 a^2 L^6 M^2 \rb^2+a^2 L^6 M \rb^3-2 a^2 L^6
   \rb^4 + 6 a^2 L^4 M^2 \rb^4+a^2 L^4 M \rb^5-6 a^2 L^4
   \rb^6+12 a^2 L^2 M^2 \rb^6 \\
&& \quad
	-\: 3 a^2 L^2 \rb^8  + 2 L^6
   M \rb^5-L^6 \rb^6+2 L^4 M \rb^7-L^4 \rb^8\big) \\
&&
	-\:2 a E L M \big(24 a^6 M^3+28 a^6 M^2
   \rb+10 a^6 M \rb^2+a^6 \rb^3
	+ 8 a^4 L^2 M^2 \rb + 8 a^4 L^2 M
   \rb^2+2 a^4 L^2 \rb^3-4 a^4 M \rb^4-2 a^4 \rb^5 \\
&& \quad
	-\: 2 a^2 L^4
   M \rb^2+a^2 L^4 \rb^3 -  2 a^2 L^2 M \rb^4-a^2 L^2 \rb^5-6
   a^2 M \rb^6-3 a^2 \rb^7+L^4 \rb^5-3 L^2 \rb^7\big)\\
&&
	+\: E^2 \left(2 a^2 M+a^2 \rb+\rb^3\right) \big(12 a^6 M^3+16 a^6 M^2 \rb+7 a^6 M 
	\rb^2+a^6 \rb^3 + 4 a^4 L^2 M^2 \rb+6 a^4 L^2 M \rb^2+2 a^4 L^2 \rb^3 \\
&& \quad
	+\: 6 a^4 M^2 \rb^3+5 a^4 M 
	\rb^4+a^4 \rb^5 - a^2 L^4 M \rb^2+a^2 L^4 \rb^3+5 a^2 L^2 M \rb^4+3 a^2 L^2 \rb^5+2 L^4 \rb^5 
	\big), \\
F^{\mathcal{K}}_{r[0]} &=& \big(8 a^8 M^3 \rb+8 a^8 M^2 \rb^2+2 a^8 M \rb^3-8 a^6 L^2 M^4+4 a^6 L^2
   M^3 \rb+12 a^6 L^2 M^2 \rb^2 + 4 a^6 L^2 M \rb^3-16 a^6 M^4
   \rb^2 \\
&& \quad
	-\: 8 a^6 M^3 \rb^3+8 a^6 M^2 \rb^4+4 a^6 M \rb^5+4 a^4
   L^4 M^3 \rb + 8 a^4 L^4 M^2 \rb^2+2 a^4 L^4 M \rb^3
	+ 8 a^4 L^2 M^3
   \rb^3+12 a^4 L^2 M^2 \rb^4 \\
&& \quad
	+\: 2 a^4 L^2 M \rb^5 - a^4 L^2
   \rb^6-8 a^4 M^3 \rb^5+2 a^4 M \rb^7+4 a^2 L^6 M^2 \rb^2+16
   a^2 L^4 M^2 \rb^4-2 a^2 L^4 \rb^6 + 4 a^2 L^2 M^2 \rb^6 \\
&& \quad
	-\: a^2 L^2
   \rb^8 +2 L^6 M \rb^5-L^6
   \rb^6+2 L^4 M \rb^7-L^4 \rb^8 \big)\\
&&
	+\: 2 a E L M
   \big(8 a^6 M^3+4 a^6 M^2 \rb-2 a^6 M \rb^2-a^6 \rb^3-4 a^4 L^2
   M^2 \rb-6 a^4 L^2 M \rb^2 - 2 a^4 L^2 \rb^3-8 a^4 M^2
   \rb^3 \\
&& \quad
	-\: 12 a^4 M \rb^4-4 a^4 \rb^5-4 a^2 L^4 M \rb^2-a^2 L^4
   \rb^3 - 10 a^2 L^2 M \rb^4-5 a^2 L^2 \rb^5-6 a^2 M \rb^6-3
   a^2 \rb^7-L^4 \rb^5-3 L^2 \rb^7\big) \\
&&
	-\: E^2 \left(2 a^2 M+a^2 \rb+\rb^3\right) \big(4 a^6
   M^3+4 a^6 M^2 \rb+a^6 M \rb^2-2 a^4 L^2 M^2 \rb - a^4 L^2 M
   \rb^2+2 a^4 M^2 \rb^3+a^4 M \rb^4 \\
&& \quad
	-\: 2 a^2 L^4 M \rb^2+a^2
   L^2 M \rb^4+a^2 L^2 \rb^5+L^4 \rb^5\big), 
\end{IEEEeqnarray*}
\begin{equation}
F_{\theta[0]} = 0, 
\end{equation}
\begin{equation}
F_{\phi[0]} = \frac{L \rbdot}{\pi  \rb \left( 2 a^2 M + a^2 \rb + L^2 \rb \right)^2 \left(\rb^2+L^2 + \frac{2 a^2 M}{\rb} + a^2 \right)^{1/2} } \left(F^{\mathcal{E}}_{\phi[0]} \mathcal{E} + F^{\mathcal{K}}_{\phi[0]} \mathcal{K} \right),
\end{equation}
where
\begin{IEEEeqnarray*}{rCl}
F^{\mathcal{E}}_{\phi[0]} &=& -2 a^4 M^2-a^4 M \rb-a^2 L^2 M \rb-4 a^2 M \rb^3-a^2
   \rb^4-L^2 \rb^4, \nonumber \\
F^{\mathcal{K}}_{\phi[0]} &=& \rb^3 \left(4 a^2 M+a^2 \rb+L^2 \rb\right).
\end{IEEEeqnarray*}
The general expressions for the higher regularization parameters, $F_{a [2]}$ and $F_{a [4]}$, are too large for
paper format and have instead been made available electronically
\cite{barrywardell.net}.  For the reader to get an understanding of the form and
size of these expressions, we include here only $F_{r [2]}$ for a circular orbit. 
This is given by
\begin{IEEEeqnarray}{rCl}
F_{r[2]} &=&  \frac{F^{\mathcal{E}}_{r[2]} \mathcal{E} + F^{\mathcal{K}}_{r[2]} \mathcal{K}}
  {6 \pi \rb^4 \left[a^2+\rb (\rb-2 M)\right]^{1/2} \left[a^4 M+2 a^3 \sqrt{M \rb^3}+a^2 \rb \left(\rb^2+M \rb-2 M^2\right)-4 a M^{3/2} \rb^{5/2}+M \rb^4\right]^3}
\end{IEEEeqnarray}
where
\begin{IEEEeqnarray*}{rCl}
F^{\mathcal{E}}_{r[2]}&=&- \Big[2 a \sqrt{M \rb}+ \rb (\rb-3 M)\Big]^{-3/2} \Big[\rb^{3/2}+a \sqrt{M}\Big] \Big[
   24 M^{9/2} a^{15}-24 M^3 \sqrt{\rb} (M^2-6 \rb M-\rb^2) a^{14} \\
&& \quad -\: 4 M^{5/2} \rb (47 M^3+13 \rb M^2-81 \rb^2 M-30\rb^3) a^{13}+2 M^2 \rb^{3/2} (93 M^4-608 \rb M^3-63 \rb^2 M^2+207 \rb^3 M  \\
&& \quad +\:123 \rb^4) a^{12}+2 M^{3/2} \rb^2 (277 M^5+285 \rb M^4-1299 \rb^2 M^3-395 \rb^3 M^2+192 \rb^4 M+132 \rb^5) a^{11} \\
&& \quad +\:M \rb^{5/2} (-543 M^6+3844 \rb M^5+495 \rb^2 M^4-2490 \rb^3 M^3-1686 \rb^4 M^2+288 \rb^5 M+156 \rb^6) a^{10} \\
&& \quad -\:2 \sqrt{M} \rb^3 (364 M^7+1087 \rb M^6-4053 \rb^2 M^5-1045 \rb^3 M^4+756 \rb^4 M^3+680 \rb^5 M^2-87 \rb^6 M-24 \rb^7) a^9 \\
&& \quad +\:\rb^{7/2} (708 M^8-5398 \rb M^7-2100 \rb^2 M^6+5292 \rb^3 M^5+4900 \rb^4 M^4-1651 \rb^5 M^3-366 \rb^6 M^2+81 \rb^7 M \\
&& \quad +\:6 \rb^8) a^8+2 \sqrt{M} \rb^4 (180 M^8+1764 \rb M^7-5867 \rb^2 M^6-1093 \rb^3 M^5-5 \rb^4 M^4+2161 \rb^5 M^3-664 \rb^6 M^2 \\
&& \quad +\:12 \rb^7 M+12 \rb^8) a^7+\rb^{9/2} (-348 M^9+2844 \rb M^8+4740 \rb^2 M^7-5611 \rb^3 M^6-5631 \rb^4 M^5+1421 \rb^5 M^4 \\
&& \quad +\:732 \rb^6 M^3-357 \rb^7 M^2+15 \rb^8 M+3 \rb^9) a^6+2 M^{3/2} \rb^6 (-1044 M^7+3324 \rb M^6+471 \rb^2 M^5+2662 \rb^3 M^4 \\
&& \quad -\:3751 \rb^4 M^3+1717 \rb^5 M^2-202 \rb^6 M+9 \rb^7) a^5+M \rb^{15/2} (-3798 M^7+3501 \rb M^6-786 \rb^2 M^5+4559 \rb^3 M^4 \\
&& \quad -\:3846 \rb^4 M^3+913 \rb^5 M^2-120 \rb^6 M+9 \rb^7) a^4-4 M^{5/2} \rb^9 (318 M^5+834 \rb M^4-618 \rb^2 M^3+176 \rb^3 M^2 \\
&& \quad -\:15 \rb^4 M-11 \rb^5) a^3+3 M^2 \rb^{21/2} (726 M^5-1417 \rb M^4+1227 \rb^2 M^3-356 \rb^3 M^2+9 \rb^4 M+3 \rb^5) a^2 \\
&& \quad +\:6 M^{7/2} \rb^{12} (222 M^3-315 \rb M^2+124 \rb^2 M-13 \rb^3) a+3 M^3 \rb^{27/2} (-20 M^3+31 \rb M^2-12 \rb^2 M+\rb^3)
   \Big], \\
F^{\mathcal{K}}_{r[2]} &=&\rb^3 \Big[2 a \sqrt{M \rb}+ \rb (\rb-3 M)\Big]^{-1/2} \Big[\rb^{3/2}+a \sqrt{M}\Big]^{-1} \Big[
   12 a^{13} M^{9/2}-12 a^{12} M^3 \sqrt{\rb} (M^2-6 M \rb-\rb^2) \\
&& \quad +\:2 a^{11} M^{5/2} \rb (-35 M^3-19 M^2 \rb+72 M \rb^2+30 \rb^3)+a^{10} M^2 \rb^{3/2} (69 M^4-452 M^3 \rb-90 M^2 \rb^2+186 M \rb^3 \\
&& \quad +\:123 \rb^4)+2 a^9 M^{3/2} \rb^2 (73 M^5+126 M^4 \rb-416 M^3 \rb^2-295 M^2 \rb^3+108 M \rb^4+66 \rb^5)+a^8 M \rb^{5/2} (-147 M^6 \\
&& \quad +\:1048 M^5 \rb+232 M^4 \rb^2-347 M^3 \rb^3-1263 M^2 \rb^4+171 M \rb^5+78 \rb^6)+2 a^7 \sqrt{M} \rb^3 (-54 M^7-342 M^6 \rb \\
&& \quad +\:989 M^5 \rb^2+625 M^4 \rb^3+100 M^3 \rb^4-499 M^2 \rb^5+33 M \rb^6+12 \rb^7)+a^6 \rb^{7/2} (114 M^8-879 M^7 \rb-639 M^6 \rb^2 \\
&& \quad -\:368 M^5 \rb^3+3058 M^4 \rb^4-158 M^3 \rb^5-252 M^2 \rb^6+9 M \rb^7+3 \rb^8)+2 a^5 M^{3/2} \rb^5 (342 M^6-981 M^5 \rb-12 M^4 \rb^2 \\
&& \quad -\:1588 M^3 \rb^3+1273 M^2 \rb^4-177 M \rb^5+9 \rb^6)+a^4 M \rb^{13/2} (1143 M^6-189 M^5 \rb-528 M^4 \rb^2-1805 M^3 \rb^3 \\
&& \quad +\:688 M^2 \rb^4-102 M \rb^5+9 \rb^6)+2 a^3 M^{5/2} \rb^8 (6 M^4+1290 M^3 \rb-543 M^2 \rb^2+73 M \rb^3+22 \rb^4) \\
&& \quad +\:3 a^2 M^2 \rb^{19/2} (-402 M^4+615 M^3 \rb-295 M^2 \rb^2+19 M \rb^3+3 \rb^4)-6 a M^{7/2} \rb^{11} (102 M^2-81 M \rb+13 \rb^2) \\
&& \quad +\:3 M^3 \rb^{25/2} (11 M^2-8 M \rb+\rb^2)
   \Big].
\end{IEEEeqnarray*}

% -------------------------------------------------------------------------------
% ------------------------------- Electromagnetic Case -----------------------------
% -------------------------------------------------------------------------------

\subsection{Electromagnetic \texorpdfstring{$\ell$}{l}-mode regularization parameters}
In the electromagnetic case, the singular part of the self-force is given by
\begin{equation} \label{eqn:SelfForceEM}
F_a = e \, A^{\rm{\sing}}_{[b, a]} u^b ,
\end{equation}
where $e$ is the charge of the particle, $u^c$ is
the four-velocity and $A^{\rm{\sing}}_{c, b}$ is the
partial derivative of the electromagnetic potential. Here, an ambiguity arises in the definition
of $u^a$ in the angular directions \emph{away} from the world line. In
Eq.~\eqref{eqn:SelfForceEM}, one is free to define $u^a(x)$ as one wishes provided
$\lim_{x \rightarrow \xb} u^a(x) = u^\ab$. A natural covariant choice would be
to define this through parallel transport, $u^a(x) = g^a{}_\bb u^\bb$. However,
in reality, it is more practical in numerical calculations to define $u^a$ such
that its components in \BL coordinates are equal to the components of $u^\ab$ in
\BL coordinates \cite{Barack:2010tm}. In doing so, the regularization parameters produced are
\begin{gather}
F_{t\lnpow{1}} =-\frac{\rb \rbdot \sgn(\Delta r)}{2[\rb \left(a^2+L^2\right)+2
   a^2 M+\rb^3]}, \nonumber \\
F_{r\lnpow{1}} = \frac{\sgn(\Delta r) \left[E \rb \left(a^2+\rb^2\right)+2 a
   M (a E-L)\right]}{2\left[a^2-2 M \rb+\rb^2\right]
   \left[\rb \left(a^2+L^2\right)+2 a^2 M+\rb^3\right]}, \nonumber \\
F_{\theta\lnpow{1}} = 0, \quad
F_{\phi\lnpow{1}} = 0,
\end{gather}
\begin{equation}
F_{t[0]} = \frac{\rbdot}{\pi  \rb^2 \left(\rb^2+L^2 + \frac{2 a^2 M}{\rb} + a^2 \right)^{3/2} \left( 2 a^2 M + a^2 \rb +L^2 \rb \right)^2} \left(F^{\mathcal{E}}_{t[0]} \mathcal{E} + F^{\mathcal{K}}_{t[0]} \mathcal{K} \right),
\end{equation}
where
\begin{IEEEeqnarray*}{rCl}
F^{\mathcal{E}}_{t[0]} &=& 
	-4 a L M \left(4 a^4 M^2+2 a^4 M \rb+2 a^2 L^2 M
   \rb-a^2 M \rb^3-a^2 \rb^4-L^2 \rb^4\right) \\
&&
	+\: E \big(-12 a^6 M^3-16 a^6 M^2 \rb-7 a^6 M \rb^2-a^6
   \rb^3-28 a^4 L^2 M^2 \rb-22 a^4 L^2 M \rb^2 - 4 a^4 L^2
   \rb^3-6 a^4 M^2 \rb^3 \\
&& \quad
	-\: 5 a^4 M \rb^4-a^4 \rb^5-15 a^2 L^4 M
   \rb^2-5 a^2 L^4 \rb^3 - 5 a^2 L^2 M \rb^4-a^2 L^2 \rb^5-2 L^6
   \rb^3\big), \\
F^{\mathcal{K}}_{t[0]} &=& 2 a L M \big(2 a^4 M^2-a^4 M \rb-a^4 \rb^2-a^2 L^2 M \rb-2 a^2 
	L^2\rb^2-2 a^2 M \rb^3-2 a^2 \rb^4 - L^4 \rb^2-2 L^2 \rb^4\big) \\
&&
	+\: E \big(4 a^6 M^3+4 a^6 M^2 \rb+a^6 M
   \rb^2+10 a^4 L^2 M^2 \rb+5 a^4 L^2 M \rb^2+2 a^4 M^2
   \rb^3 + a^4 M \rb^4+4 a^2 L^4 M \rb^2 \\
&& \quad
	+\: a^2 L^2 M \rb^4-a^2 L^2
   \rb^5-L^4 \rb^5\big),
\end{IEEEeqnarray*}
\begin{equation}
F_{r[0]} = \frac{\left(F^{\mathcal{E}}_{r[0]} \mathcal{E} + F^{\mathcal{K}}_{r[0]} \mathcal{K} \right)}{\pi  \rb^3 \left(\rb^2+L^2 + \frac{2 a^2 M}{\rb} + a^2 \right)^{3/2} \left( 2 a^2 M + a^2 \rb + L^2 \rb \right)^2 \left( a^2 - 2 M \rb + \rb^2\right)} ,
\end{equation}
where
\begin{IEEEeqnarray*}{rCl}
F^{\mathcal{E}}_{r[0]} &=& 
L^2 \big(24 a^6 M^4+28 a^6 M^3 \rb-6 a^6 M^2 \rb^2-11 a^6 M \rb^3-2
   a^6 \rb^4+56 a^4 L^2 M^3 \rb + 24 a^4 L^2 M^2 \rb^2-18 a^4 L^2 M
   \rb^3 \\
&& \quad
	-\: 6 a^4 L^2 \rb^4+52 a^4 M^3 \rb^3+20 a^4 M^2 \rb^4 - 11 a^4 M \rb^5-3 a^4 \rb^6+30 a^2 L^4 M^2 \rb^2-3 a^2 L^4 M
   \rb^3-6 a^2 L^4 \rb^4 \\
&& \quad
	+\: 42 a^2 L^2 M^2 \rb^4 - 5 a^2 L^2 M
   \rb^5-6 a^2 L^2 \rb^6+8 a^2 M^2 \rb^6-2 a^2 M \rb^7-a^2
   \rb^8+4 L^6 M \rb^3-2 L^6 \rb^4 + 6 L^4 M \rb^5 \\
&& \quad
	-\: 3 L^4
   \rb^6+2 L^2 M \rb^7-L^2 \rb^8\big) \\
&& 
	-\: 2 a E L M \big(24
   a^6 M^3+36 a^6 M^2 \rb+18 a^6 M \rb^2+3 a^6 \rb^3+56 a^4 L^2 M^2
   \rb + 48 a^4 L^2 M \rb^2+10 a^4 L^2 \rb^3+24 a^4 M^2 \rb^3 \\
&& \quad
	+\: 24
   a^4 M \rb^4+6 a^4 \rb^5+30 a^2 L^4 M \rb^2 + 11 a^2 L^4
   \rb^3+22 a^2 L^2 M \rb^4+9 a^2 L^2 \rb^5+6 a^2 M \rb^6+3
   a^2 \rb^7+4 L^6 \rb^3 \\
&& \quad
	+\: 3 L^4 \rb^5 + 3 L^2 \rb^7\big) \\
&&
	+\: E^2 \left(2 a^2 M+a^2 \rb+\rb^3\right) \big(12 a^6 M^3+16 a^6 M^2
   \rb+7 a^6 M \rb^2+a^6 \rb^3 + 28 a^4 L^2 M^2 \rb+22 a^4 L^2 M
   \rb^2+4 a^4 L^2 \rb^3 \\
&& \quad
	+\: 6 a^4 M^2 \rb^3+5 a^4 M \rb^4+a^4
   \rb^5 + 15 a^2 L^4 M \rb^2+5 a^2 L^4 \rb^3+5 a^2 L^2 M
   \rb^4+a^2 L^2 \rb^5+2 L^6 \rb^3\big), \\
F^{\mathcal{K}}_{r[0]} &=& 
	-L^2 \big(8 a^6 M^4+12 a^6 M^3
   \rb-2 a^6 M \rb^3+20 a^4 L^2 M^3 \rb+8 a^4 L^2 M^2 \rb^2 - 4 a^4 L^2 M \rb^3+32 a^4 M^3 \rb^3+12 a^4 M^2 \rb^4 \\
&& \quad
	-\: 6 a^4 M
   \rb^5-a^4 \rb^6+8 a^2 L^4 M^2 \rb^2 - 2 a^2 L^4 M \rb^3+24
   a^2 L^2 M^2 \rb^4-4 a^2 L^2 M \rb^5-2 a^2 L^2 \rb^6+8 a^2 M^2
   \rb^6 \\
&& \quad
	-\: 2 a^2 M \rb^7 - a^2 \rb^8+2 L^4 M \rb^5-L^4
   \rb^6+2 L^2 M \rb^7-L^2 \rb^8\big)\\
&&
	+\: 2 a E L M \big(8
   a^6 M^3+12 a^6 M^2 \rb+6 a^6 M \rb^2+a^6 \rb^3+20 a^4 L^2 M^2
   \rb + 14 a^4 L^2 M \rb^2+2 a^4 L^2 \rb^3+16 a^4 M^2 \rb^3 \\
&& \quad
	+\: 16
   a^4 M \rb^4+4 a^4 \rb^5+8 a^2 L^4 M \rb^2 + a^2 L^4 \rb^3+14
   a^2 L^2 M \rb^4+5 a^2 L^2 \rb^5+6 a^2 M \rb^6+3 a^2
   \rb^7+L^4 \rb^5+3 L^2 \rb^7\big) \\
&&
	-\: E^2 \left(2 a^2 M+a^2 \rb+\rb^3\right) \big(4 a^6
   M^3+4 a^6 M^2 \rb+a^6 M \rb^2+10 a^4 L^2 M^2 \rb + 5 a^4 L^2 M
   \rb^2+2 a^4 M^2 \rb^3+a^4 M \rb^4 \\
&& \quad
	+\: 4 a^2 L^4 M \rb^2+a^2 L^2
   M \rb^4-a^2 L^2 \rb^5 - L^4 \rb^5\big),
\end{IEEEeqnarray*}
\begin{equation}
F_{\theta[0]}  = 0,
\end{equation}
\begin{equation}
F_{\phi[0]} = \frac{ L \rbdot \left(F^{\mathcal{E}}_{\phi[0]} \mathcal{E} + F^{\mathcal{K}}_{ \phi[0]} \mathcal{K} \right)}{\pi  \rb \left(\rb^2+L^2 + \frac{2 a^2 M}{\rb} + a^2 \right)^{1/2} \left( 2 a^2 M + a^2 \rb + L^2 \rb \right)^2} ,
\end{equation}
where
\begin{IEEEeqnarray*}{rCl}
F^{\mathcal{E}}_{\phi[0]} &=& 14 a^4 M^2+11 a^4 M \rb+2 a^4 \rb^2+11 a^2 L^2 M \rb+4 a^2 L^2
   \rb^2+4 a^2 M \rb^3+a^2 \rb^4 + 2 L^4 \rb^2+L^2 \rb^4, \\
F^{\mathcal{K}}_{\phi[0]} &=& -4 a^4 M^2-2 a^4 M \rb-2 a^2 L^2 M \rb-4 a^2 M \rb^3-a^2
   \rb^4-L^2 \rb^4.
\end{IEEEeqnarray*}
As with the scalar case, $F_{a[2]}$ proves too large to include in paper format
and so is available electronically \cite{barrywardell.net}; again we provide
$F_{r[2]}$ for circular orbits below to allow the reader to get an understanding
of the structure of the parameters:
\begin{IEEEeqnarray}{rCl}
F_{r[2]} &=&  \frac{F^{\mathcal{E}}_{r[2]} \mathcal{E} + F^{\mathcal{K}}_{r[2]} \mathcal{K}}
  {6 \pi \rb^4  [a^2+\rb (\rb-2 M) ]^{1/2}  [a^4 M+2 a^3 \sqrt{M \rb^3}+a^2 \rb  (\rb^2+M \rb-2 M^2 )-4 a M^{3/2} \rb^{5/2}+M \rb^4 ]^3}
\end{IEEEeqnarray}
where
\begin{IEEEeqnarray*}{rCl}
F^{\mathcal{E}}_{r[2]}&=&- \Big[2 a \sqrt{M \rb}+ \rb (\rb-3 M)\Big]^{-3/2} \Big[\rb^{3/2}+a \sqrt{M}\Big] \Big[
  48 M^{9/2} a^{15}-24 M^3 \sqrt{\rb} (2 M^2-12 \rb M+\rb^2) a^{14} \\
&& \quad -\:4 M^{5/2} \rb (91 M^3+29 \rb M^2-135 \rb^2 M+30 \rb^3) a^{13}+2 M^2 \rb^{3/2} (177 M^4-1168 \rb M^3+351 \rb^2 M^2+27 \rb^3 M \\
&& \quad -\:123 \rb^4) a^{12}+2 M^{3/2} \rb^2 (515 M^5+579 \rb M^4-2265 \rb^2 M^3+1343 \rb^3 M^2-516 \rb^4 M-132 \rb^5) a^{11} \\
&& \quad -\:M \rb^{5/2} (969 M^6-6992 \rb M^5+3435 \rb^2 M^4-654 \rb^3 M^3-3426 \rb^4 M^2+1368 \rb^5 M+156 \rb^6) a^{10} \\
&& \quad -\:2 \sqrt{M} \rb^3 (644 M^7+2009 \rb M^6-7611 \rb^2 M^5+8377 \rb^3 M^4-4548 \rb^4 M^3-788 \rb^5 M^2+375 \rb^6 M+24 \rb^7) a^9 \\
&& \quad -\:\rb^{7/2} (-1164 M^8+9146 \rb M^7-4464 \rb^2 M^6-1368 \rb^3 M^5+16348 \rb^4 M^4-8911 \rb^5 M^3+330 \rb^6 M^2+189 \rb^7 M \\
&& \quad +\:6 \rb^8) a^8-2 \sqrt{M} \rb^4 (-300 M^8-2940 \rb M^7+11773 \rb^2 M^6-19669 \rb^3 M^5+13183 \rb^4 M^4+1765 \rb^5 M^3 \\
&& \quad -\:1648 \rb^6 M^2+264 \rb^7 M+12 \rb^8) a^7-\rb^{9/2} (516 M^9-4404 \rb M^8-2436 \rb^2 M^7+16697 \rb^3 M^6-39699 \rb^4 M^5 \\
&& \quad +\:16229 \rb^5 M^4-3516 \rb^6 M^3-609 \rb^7 M^2+123 \rb^8 M+3 \rb^9) a^6-2 M^{3/2} \rb^6 (1548 M^7-6900 \rb M^6+14991 \rb^2 M^5 \\
&& \quad -\:10586 \rb^3 M^4+1325 \rb^4 M^3+2941 \rb^5 M^2-322 \rb^6 M-27 \rb^7) a^5-M \rb^{15/2} (5706 M^7-21603 \rb M^6 \\
&& \quad +\:39138 \rb^2 M^5-14389 \rb^3 M^4-1578 \rb^4 M^3-599 \rb^5 M^2+228 \rb^6 M+9 \rb^7) a^4-4 M^{5/2} \rb^9 (546 M^5-1902 \rb M^4 \\
&& \quad -\:2394 \rb^2 M^3+1876 \rb^3 M^2+93 \rb^4 M-97 \rb^5) a^3+3 M^2 \rb^{21/2} (786 M^5-3755 \rb M^4+1489 \rb^2 M^3+360 \rb^3 M^2 \\
&& \quad -\:133 \rb^4 M-3 \rb^5) a^2+6 M^{7/2} \rb^{12} (138 M^3+343 \rb M^2-348 \rb^2 M+73 \rb^3) a-3 M^3 \rb^{27/2} (196 M^3-165 \rb M^2 \\
&& \quad +\:32 \rb^2 M+\rb^3)
   \Big], \\
F^{\mathcal{K}}_{r [2]} &=&-\rb^{5/2} \Big[2 a \sqrt{M \rb}+ \rb (\rb-3 M)\Big]^{-1/2} \Big[\rb^{3/2}+a \sqrt{M}\Big]^{-1} \Big[
  144 a^{14} M^4+12 a^{13} M^{7/2} \sqrt{\rb} (72 \rb-13 M) \\
&& \quad -\:12 a^{12} M^3 \rb (73 M^2+40 M \rb-181 \rb^2)+2 a^{11} M^{5/2} \rb^{3/2} (493 M^3-2887 M^2 \rb-180 M \rb^2+1470 \rb^3) \\
&& \quad +\:a^{10} M^2 \rb^2 (1725 M^4+4108 M^3 \rb-13842 M^2 \rb^2+426 M \rb^3+2283 \rb^4)+2 a^9 M^{3/2} \rb^{5/2} (-1031 M^5+6246 M^4 \rb \\
&& \quad +\:2452 M^3 \rb^2-8011 M^2 \rb^3+528 M \rb^4+498 \rb^5)+a^8 M \rb^3 (-1011 M^6-10712 M^5 \rb+30400 M^4 \rb^2+517 M^3 \rb^3 \\
&& \quad -\:9303 M^2 \rb^4+963 M \rb^5+222 \rb^6)+2 a^7 \sqrt{M} \rb^{7/2} (714 M^7-4134 M^6 \rb-8611 M^5 \rb^2+15349 M^4 \rb^3 \\
&& \quad -\:1664 M^3 \rb^4-1183 M^2 \rb^5+237 M \rb^6+12 \rb^7)+a^6 \rb^4 (-174 M^8+8817 M^7 \rb-21807 M^6 \rb^2-7640 M^5 \rb^3 \\
&& \quad +\:12274 M^4 \rb^4-3014 M^3 \rb^5-204 M^2 \rb^6+105 M \rb^7+3 \rb^8)-2 a^5 M^{3/2} \rb^{11/2} (522 M^6-8811 M^5 \rb \\
&& \quad +\:10548 M^4 \rb^2-1328 M^3 \rb^3-805 M^2 \rb^4+369 M \rb^5+27 \rb^6)+a^4 M \rb^7 (-1881 M^6+10731 M^5 \rb-3648 M^4 \rb^2 \\
&& \quad +\:763 M^3 \rb^3-32 M^2 \rb^4+210 M \rb^5+9 \rb^6)-2 a^3 M^{5/2} \rb^{17/2} (282 M^4+1470 M^3 \rb-1797 M^2 \rb^2+203 M \rb^3 \\
&& \quad +\:194 \rb^4)+3 a^2 M^2 \rb^{10} (318 M^4-921 M^3 \rb-23 M^2 \rb^2+131 M \rb^3+3 \rb^4)+6 a M^{7/2} \rb^{23/2} (42 M^2+169 M \rb \\
&& \quad -\:73 \rb^2)+3 M^3 \rb^{13} (-85 M^2+32 M \rb+\rb^2)
   \Big].
\end{IEEEeqnarray*}

% -------------------------------------------------------------------------------
% ---------------------------- Gravitational Case -----------------------------------
% -------------------------------------------------------------------------------

\subsection{Gravitational \texorpdfstring{$\ell$}{l}-mode regularization parameters}

% ---------------------------- Self-force regularization -----------------------------

\subsubsection{Self-force regularization}
The singular part of the self-force on a point mass is given by
\begin{equation}
F^a = k^{abcd} \hb^{\rm \sing}_{bc;d},
\end{equation}
where
\begin{equation}
k^{abcd} \equiv \frac12 g^{ad} u^b u^c - g^{ab} u^c u^d - 
  \frac12 u^a u^b u^c u^d + \frac14 u^a g^{b c} u^d + 
 \frac14 g^{a d} g^{b c},
\end{equation}
and $\hb^{\rm \sing}_{bc}$ is the trace-reversed singular metric perturbation.  Note that, as
in the electromagnetic case, an ambiguity arises here due to the presence of
terms involving the four-velocity at $x$. One is free to arbitrarily choose how
to define this provided $\lim_{x\to\bar{x}} u^a = u^{\ab}$. Following Barack and
Sago~\cite{Barack:2010tm}, we choose to take the Boyer-Lindquist components of the
four-velocity at $x$ to be exactly those at $\xb$.  The regularization parameters in
the gravitational case are given by
\begin{gather}
F^{t}{}_{\lnpow{1}} = -\frac{[\rb^3 + a^2 (2 M + \rb)]  \rbdot \sgn(\Delta r)}{2[a^2-2M\rb + \rb^2][\rb
   \left(a^2+L^2\right)+2 a^2 M+\rb^3]} , \\
F^{r}{}_{\lnpow{1}} = -\frac{\sgn(\Delta r) \left[E \rb
   \left(a^2+\rb^2\right)+2 a M (a
   E-L)\right]}{2 \rb^2
   \left[\rb \left(a^2+L^2\right)+2 a^2 M+\rb^3\right]}, \\
F^{\theta}{}_{ \lnpow{1}} = 0,  \qquad
F^{\phi}{}_{ \lnpow{1}} =-\frac{a M \rbdot \sgn(\Delta r)}{[a^2-2M\rb + \rb^2][\rb
   \left(a^2+L^2\right)+2 a^2 M+\rb^3]},
\end{gather}
\begin{equation}
F^{t}{}_{[0]} = \frac{\rbdot \left(F^{t}_{\mathcal{E}}{}_{[0]} \mathcal{E} + F^{t}_{\mathcal{K}}{}_{ [0]} \mathcal{K} \right)}{\pi  \rb^3 [a^2-2M\rb + \rb^2]\left[\rb^2+L^2 + \frac{2 a^2 M}{\rb} + a^2 \right]^{3/2} \left[ 2 a^2 M + a^2 \rb + L^2 \rb \right]^2} ,
\end{equation}
where
\begin{IEEEeqnarray*}{rCl}
F^{t}_{\mathcal{E}}{}_{[0]} &=& 
  2 a L M (12 a^6 M^3 + 20 a^6 M^2 \rb + 28 a^4 L^2 M^2 \rb + 
     11 a^6 M \rb^2 + 26 a^4 L^2 M \rb^2 + 15 a^2 L^4 M \rb^2 + 2 a^6 \rb^3 + 
     6 a^4 L^2 \rb^3 \\
     && \quad
     +\: 6 a^2 L^4 \rb^3 + 2 L^6 \rb^3 + 18 a^4 M^2 \rb^3 + 
     19 a^4 M \rb^4 + 17 a^2 L^2 M \rb^4 + 5 a^4 \rb^5 + 8 a^2 L^2 \rb^5 + 
     3 L^4 \rb^5 + 6 a^2 M \rb^6 \\
     && \quad
     +\: 3 a^2 \rb^7 + 3 L^2 \rb^7) - E (2 a^2 M + 
     a^2 \rb + \rb^3) (12 a^6 M^3 + 16 a^6 M^2 \rb + 28 a^4 L^2 M^2 \rb + 
     7 a^6 M \rb^2 + 22 a^4 L^2 M \rb^2 \\
     && \quad
     +\: 15 a^2 L^4 M \rb^2 + a^6 \rb^3 + 
     4 a^4 L^2 \rb^3 + 5 a^2 L^4 \rb^3 + 2 L^6 \rb^3 + 6 a^4 M^2 \rb^3 + 
     5 a^4 M \rb^4 + 5 a^2 L^2 M \rb^4 + a^4 \rb^5 + 
     a^2 L^2 \rb^5), \\
F^t_{\mathcal{K}}{}_{[0]} &=&
 -2 a L M (4 a^6 M^3 + 8 a^6 M^2 \rb + 10 a^4 L^2 M^2 \rb + 5 a^6 M \rb^2 + 
     9 a^4 L^2 M \rb^2 + 4 a^2 L^4 M \rb^2 + a^6 \rb^3 + 2 a^4 L^2 \rb^3 + 
     a^2 L^4 \rb^3 \\
     && \quad
     +\: 14 a^4 M^2 \rb^3 + 15 a^4 M \rb^4 + 13 a^2 L^2 M \rb^4 + 
     4 a^4 \rb^5 + 6 a^2 L^2 \rb^5 + 2 L^4 \rb^5 + 6 a^2 M \rb^6 + 
     3 a^2 \rb^7 + 3 L^2 \rb^7) \\
     && \quad
     +\: E (2 a^2 M + a^2 \rb + \rb^3) (4 a^6 M^3 + 
     4 a^6 M^2 \rb + 10 a^4 L^2 M^2 \rb + a^6 M \rb^2 + 5 a^4 L^2 M \rb^2 + 
     4 a^2 L^4 M \rb^2 + 2 a^4 M^2 \rb^3 \\
     && \quad
     +\: a^4 M \rb^4 + a^2 L^2 M \rb^4 - 
     a^2 L^2 \rb^5 - L^4 \rb^5),
\end{IEEEeqnarray*}
\begin{equation}
F^r{}_{[0]} = \frac{ \left(F^r_{\mathcal{E}}{}_{[0]} \mathcal{E} + F^r_{\mathcal{K}}{}_{[0]} \mathcal{K} \right)}{\pi  \rb^5 \left(\rb^2+L^2 + \frac{2 a^2 M}{\rb} + a^2 \right)^{3/2} \left( 2 a^2 M + a^2 \rb + L^2 \rb \right)^2} ,
\end{equation}
where
\begin{IEEEeqnarray*}{rCl}
F^r_{\mathcal{E}}{}_{[0]} &=& 
-L^2 (24 a^6 M^4 + 28 a^6 M^3 \rb + 56 a^4 L^2 M^3 \rb - 6 a^6 M^2 \rb^2 + 
     24 a^4 L^2 M^2 \rb^2 + 30 a^2 L^4 M^2 \rb^2 - 11 a^6 M \rb^3 \\
     && \quad
     -\: 18 a^4 L^2 M \rb^3 - 3 a^2 L^4 M \rb^3 + 4 L^6 M \rb^3 + 
     52 a^4 M^3 \rb^3 - 2 a^6 \rb^4 - 6 a^4 L^2 \rb^4 - 6 a^2 L^4 \rb^4 - 
     2 L^6 \rb^4 + 20 a^4 M^2 \rb^4 \\
     && \quad
     +\: 42 a^2 L^2 M^2 \rb^4 - 11 a^4 M \rb^5 - 
     5 a^2 L^2 M \rb^5 + 6 L^4 M \rb^5 - 3 a^4 \rb^6 - 6 a^2 L^2 \rb^6 - 
     3 L^4 \rb^6 + 8 a^2 M^2 \rb^6 - 2 a^2 M \rb^7 \\
     && \quad
     +\: 2 L^2 M \rb^7 - a^2 \rb^8 -
      L^2 \rb^8) + 
  2 a E L M(24 a^6 M^3 + 36 a^6 M^2 \rb + 56 a^4 L^2 M^2 \rb + 
     18 a^6 M \rb^2 + 48 a^4 L^2 M \rb^2 \\
     && \quad
     +\: 30 a^2 L^4 M \rb^2 + 3 a^6 \rb^3 + 
     10 a^4 L^2 \rb^3 + 11 a^2 L^4 \rb^3 + 4 L^6 \rb^3 + 24 a^4 M^2 \rb^3 + 
     24 a^4 M \rb^4 + 22 a^2 L^2 M \rb^4 + 6 a^4 \rb^5 \\
     && \quad
     +\: 9 a^2 L^2 \rb^5 + 
     3 L^4 \rb^5 + 6 a^2 M \rb^6 + 3 a^2 \rb^7 + 
     3 L^2 \rb^7)-  E^2 (2 a^2 M + a^2 \rb + 
     \rb^3) (12 a^6 M^3 + 16 a^6 M^2 \rb \\
     && \quad
     +\: 28 a^4 L^2 M^2 \rb + 
     7 a^6 M \rb^2 + 22 a^4 L^2 M \rb^2 + 15 a^2 L^4 M \rb^2 + a^6 \rb^3 + 
     4 a^4 L^2 \rb^3 + 5 a^2 L^4 \rb^3 + 2 L^6 \rb^3 + 6 a^4 M^2 \rb^3 \\
     && \quad
     +\: 5 a^4 M \rb^4 + 5 a^2 L^2 M \rb^4 + a^4 \rb^5 + 
     a^2 L^2 \rb^5), \\
F^r_{\mathcal{K}}{}_{[0]} &=& 
L^2 (8 a^6 M^4 + 12 a^6 M^3 \rb + 20 a^4 L^2 M^3 \rb + 
     8 a^4 L^2 M^2 \rb^2 + 8 a^2 L^4 M^2 \rb^2 - 2 a^6 M \rb^3 - 
     4 a^4 L^2 M \rb^3 - 2 a^2 L^4 M \rb^3 \\
     && \quad
     +\: 32 a^4 M^3 \rb^3 + 
     12 a^4 M^2 \rb^4 + 24 a^2 L^2 M^2 \rb^4 - 6 a^4 M \rb^5 - 
     4 a^2 L^2 M \rb^5 + 2 L^4 M \rb^5 - a^4 \rb^6 - 2 a^2 L^2 \rb^6 - 
     L^4 \rb^6 \\
     && \quad
     +\: 8 a^2 M^2 \rb^6 - 2 a^2 M \rb^7 + 2 L^2 M \rb^7 - a^2 \rb^8 - 
     L^2 \rb^8) - 
  2 a E L M (8 a^6 M^3 + 12 a^6 M^2 \rb + 20 a^4 L^2 M^2 \rb + 
     6 a^6 M \rb^2 \\
     && \quad
     +\: 14 a^4 L^2 M \rb^2 + 8 a^2 L^4 M \rb^2 + a^6 \rb^3 + 
     2 a^4 L^2 \rb^3 + a^2 L^4 \rb^3 + 16 a^4 M^2 \rb^3 + 16 a^4 M \rb^4 + 
     14 a^2 L^2 M \rb^4 + 4 a^4 \rb^5 \\
     && \quad
     +\: 5 a^2 L^2 \rb^5 + L^4 \rb^5 + 
     6 a^2 M \rb^6 + 3 a^2 \rb^7 + 
     3 L^2 \rb^7) + E^2 (2 a^2 M + a^2 \rb + 
     \rb^3) (4 a^6 M^3 + 4 a^6 M^2 \rb + 10 a^4 L^2 M^2 \rb \\
     && \quad
     +\: a^6 M \rb^2 + 
     5 a^4 L^2 M \rb^2 + 4 a^2 L^4 M \rb^2 + 2 a^4 M^2 \rb^3 + a^4 M \rb^4 + 
     a^2 L^2 M \rb^4 - a^2 L^2 \rb^5 - L^4 \rb^5) ,
\end{IEEEeqnarray*}
\begin{equation}
F^{\theta}{}_{[0]}  = 0,
\end{equation}
\begin{equation}
F^{\phi}{}_{[0]} = \frac{ \rbdot \left(F^{\phi}_{\mathcal{E}}{}_{[0]} \mathcal{E} + F^{\phi}_{\mathcal{K}}{}_{[0]} \mathcal{K} \right)}{\pi  \rb^3 \left(\rb^2+L^2 + \frac{2 a^2 M}{\rb} + a^2 \right)^{3/2} \left( 2 a^2 M + a^2 \rb + L^2 \rb \right)^2 \left( a^2 - 2 M \rb  + \rb^2 \right)} ,
\end{equation}
where
\begin{IEEEeqnarray*}{rCl}
F^{\phi}_{\mathcal{E}}{}_{[0]} &=&
L (24 a^6 M^4 + 28 a^6 M^3 \rb + 56 a^4 L^2 M^3 \rb - 6 a^6 M^2 \rb^2 + 
    24 a^4 L^2 M^2 \rb^2 + 30 a^2 L^4 M^2 \rb^2 - 11 a^6 M \rb^3 - 
    18 a^4 L^2 M \rb^3 \\
    && \quad
    -\: 3 a^2 L^4 M \rb^3 + 4 L^6 M \rb^3 + 
    52 a^4 M^3 \rb^3 - 2 a^6 \rb^4 - 6 a^4 L^2 \rb^4 - 6 a^2 L^4 \rb^4 - 
    2 L^6 \rb^4 + 20 a^4 M^2 \rb^4 + 42 a^2 L^2 M^2 \rb^4 \\
    && \quad
    -\: 11 a^4 M \rb^5 - 
    5 a^2 L^2 M \rb^5 + 6 L^4 M \rb^5 - 3 a^4 \rb^6 - 6 a^2 L^2 \rb^6 - 
    3 L^4 \rb^6 + 8 a^2 M^2 \rb^6 - 2 a^2 M \rb^7 + 2 L^2 M \rb^7 - a^2 \rb^8 \\
    && \quad
    - \: L^2 \rb^8) - 
    2 a E M (12 a^6 M^3 + 16 a^6 M^2 \rb + 28 a^4 L^2 M^2 \rb + 7 a^6 M \rb^2 + 
    22 a^4 L^2 M \rb^2 + 15 a^2 L^4 M \rb^2 + a^6 \rb^3 \\
    && \quad
    +\: 4 a^4 L^2 \rb^3 + 
    5 a^2 L^4 \rb^3 + 2 L^6 \rb^3 + 6 a^4 M^2 \rb^3 + 5 a^4 M \rb^4 + 
    5 a^2 L^2 M \rb^4 + a^4 \rb^5 + a^2 L^2 \rb^5), \\
F^{\phi}_{\mathcal{K}}{}_{[0]} &=& 
-L (8 a^6 M^4 + 12 a^6 M^3 \rb + 20 a^4 L^2 M^3 \rb + 8 a^4 L^2 M^2 \rb^2 + 
    8 a^2 L^4 M^2 \rb^2 - 2 a^6 M \rb^3 - 4 a^4 L^2 M \rb^3 - 
    2 a^2 L^4 M \rb^3 \\
    && \quad
    +\: 32 a^4 M^3 \rb^3 + 12 a^4 M^2 \rb^4 + 
    24 a^2 L^2 M^2 \rb^4 - 6 a^4 M \rb^5 - 4 a^2 L^2 M \rb^5 + 2 L^4 M \rb^5 -
     a^4 \rb^6 - 2 a^2 L^2 \rb^6 - L^4 \rb^6 \\
    && \quad
    +\: 8 a^2 M^2 \rb^6 - 2 a^2 M \rb^7 + 2 L^2 M \rb^7 - a^2 \rb^8 - L^2 \rb^8) + 
    2 a E M (4 a^6 M^3 + 4 a^6 M^2 \rb + 10 a^4 L^2 M^2 \rb + a^6 M \rb^2 \\
    && \quad
    +\: 5 a^4 L^2 M \rb^2 + 4 a^2 L^4 M \rb^2 + 2 a^4 M^2 \rb^3 + a^4 M \rb^4 + 
    a^2 L^2 M \rb^4 - a^2 L^2 \rb^5 - L^4 \rb^5).
\end{IEEEeqnarray*}
As with the scalar and electromagnetic cases, $F^a_{[2]} $ is too large for paper
format and so is available electronically \cite{barrywardell.net}. Instead, we give
here $F^r_{[2]}$ for circular orbits,
\begin{IEEEeqnarray}{rCl}
F^r_{[2]} &=&  \frac{(F^{r}_{\mathcal{E}[2]} \mathcal{E} + F^{r}_{\mathcal{K}[2]} \mathcal{K})
  \sqrt{a^2-2Mr+r^2}}
  {6\pi \rb^{13/2}  [a^4 M+2 a^3 \sqrt{M \rb^3}+a^2 \rb  (\rb^2+M \rb-2 M^2 )-4 a M^{3/2} \rb^{5/2}+M \rb^4 ]^3}
\end{IEEEeqnarray}
where
\begin{IEEEeqnarray*}{rCl}
F^{r}_{\mathcal{E}[2]}&=& \Big[2 a \sqrt{M}+ \sqrt{\rb} (\rb-3 M)\Big]^{-3/2} \Big[\rb^{3/2}+a \sqrt{M}\Big] \Big[
264 M^{9/2} a^{15}-24 M^3 \sqrt{\rb} (11 M^2-66 \rb M+5 \rb^2) a^{14}\\
&& \quad
 -\:20 M^{5/2} \rb (101 M^3+31 \rb M^2-171 \rb^2 M+30 \rb^3) a^{13}+2 M^2 \rb^{3/2} (987
   M^4-6496 \rb M^3+1299 \rb^2 M^2\\
&& \quad
 +\:1329 \rb^3 M-615 \rb^4) a^{12}+2 M^{3/2} \rb^2 (2891 M^5+3171 \rb M^4-13533 \rb^2 M^3+4859 \rb^3 M^2-576 \rb^4 M \\
&& \quad
 -\:660 \rb^5) a^{11}-M
   \rb^{5/2} (5505 M^6-39500 \rb M^5+11775 \rb^2 M^4+10254 \rb^3 M^3-12678 \rb^4 M^2+3576 \rb^5 M \\
&& \quad
 +\:780 \rb^6) a^{10}-2 \sqrt{M} \rb^3 (3668 M^7+11297 \rb M^6-43179 \rb^2
   M^5+29605 \rb^3 M^4-10932 \rb^4 M^3-3992 \rb^5 M^2 \\
&& \quad
 +\:1335 \rb^6 M+120 \rb^7) a^9+\rb^{7/2} (6780 M^8-52778 \rb M^7+10692 \rb^2 M^6+26916 \rb^3 M^5-51340 \rb^4 M^4 \\
&& \quad
 +\:23203 \rb^5
   M^3+1902 \rb^6 M^2-993 \rb^7 M-30 \rb^8) a^8+2 \sqrt{M} \rb^4 (1740 M^8+17052 \rb M^7-64693 \rb^2 M^6 \\
&& \quad
 +\:73093 \rb^3 M^5-45691 \rb^4 M^4-4945 \rb^5 M^3+3688 \rb^6 M^2-204 \rb^7
   M-108 \rb^8) a^7+\rb^{9/2} (-3108 M^9 \\
&& \quad
 +\:26148 \rb M^8+23964 \rb^2 M^7-82757 \rb^3 M^6+132831 \rb^4 M^5-43349 \rb^5 M^4+8772 \rb^6 M^3+1941 \rb^7 M^2 \\
&& \quad
 -\:207 \rb^8 M-27 \rb^9)
   a^6-2 M^{3/2} \rb^6 (9324 M^7-37572 \rb M^6+58551 \rb^2 M^5-49466 \rb^3 M^4+16841 \rb^4 M^3 \\
&& \quad
 +\:5029 \rb^5 M^2+422 \rb^6 M-327 \rb^7) a^5+M \rb^{15/2} (-34218 M^7+96915 \rb
   M^6-156318 \rb^2 M^5+59569 \rb^3 M^4 \\
&& \quad
 +\:2910 \rb^4 M^3+3935 \rb^5 M^2-1488 \rb^6 M+39 \rb^7) a^4-4 M^{5/2} \rb^9 (3138 M^5-5106 \rb M^4-12726 \rb^2 M^3 \\
&& \quad
 +\:9112 \rb^3 M^2+111 \rb^4
   M-421 \rb^5) a^3+3 M^2 \rb^{21/2} (5322 M^5-19271 \rb M^4+9205 \rb^2 M^3+876 \rb^3 M^2 \\
&& \quad
 -\:697 \rb^4 M+21 \rb^5) a^2+6 M^{7/2} \rb^{12} (1218 M^3+427 \rb M^2-1020 \rb^2
   M+253 \rb^3) a-3 M^3 \rb^{27/2} (844 M^3 \\
&& \quad
 -\:753 \rb M^2+164 \rb^2 M+\rb^3)
  \Big], \\
F^{r}_{\mathcal{K} [2]} &=& \rb^3 \Big[2 a \sqrt{M \rb}+ \rb (\rb-3 M)\Big]^{-1/2} \Big[\rb^{3/2}+a \sqrt{M}\Big]^{-1} \Big[
576 a^{14} M^4+12 a^{13} M^{7/2} \sqrt{\rb} (284 \rb-51 M)\\
&& \quad
  -\:12 a^{12} M^3 \rb (295 M^2+146 M \rb-711 \rb^2)+2 a^{11} M^{5/2} \rb^{3/2} (1981 M^3-11563 M^2 \rb-624 M
   \rb^2+5838 \rb^3)\\
&& \quad
 +\:a^{10} M^2 \rb^2 (6981 M^4+16108 M^3 \rb-54498 M^2 \rb^2+714 M \rb^3+9411 \rb^4)+2 a^9 M^{3/2} \rb^{5/2} (-4247 M^5 \\
&& \quad
 +\:25278 M^4 \rb+9304 M^3
   \rb^2-30511 M^2 \rb^3+708 M \rb^4+2250 \rb^5)+a^8 M \rb^3 (-3891 M^6-44168 M^5 \rb \\
&& \quad
 +\:121408 M^4 \rb^2+3445 M^3 \rb^3-33279 M^2 \rb^4+987 M \rb^5+1254 \rb^6)+2 a^7 \sqrt{M}
   \rb^{7/2} (3018 M^7-16374 M^6 \rb \\
&& \quad
 -\:35107 M^5 \rb^2+57481 M^4 \rb^3-3572 M^3 \rb^4-3835 M^2 \rb^5+297 M \rb^6+108 \rb^7)+a^6 \rb^4 (-1038 M^8 \\
&& \quad
 +\:37905 M^7 \rb-87903 M^6
   \rb^2-27632 M^5 \rb^3+35290 M^4 \rb^4-5726 M^3 \rb^5-1116 M^2 \rb^6+201 M \rb^7+27 \rb^8) \\
&& \quad
 -\:2 a^5 M^{3/2} \rb^{11/2} (3114 M^6-38187 M^5 \rb+41580 M^4 \rb^2-7508 M^3 \rb^3+527 M^2
   \rb^4+177 M \rb^5+327 \rb^6) \\
&& \quad
 +\:a^4 M \rb^7 (-10953 M^6+43491 M^5 \rb-8904 M^4 \rb^2+3571 M^3 \rb^3-872 M^2 \rb^4+1242 M \rb^5-39 \rb^6) \\
&& \quad
 -\:2 a^3 M^{5/2} \rb^{17/2} (1146
   M^4+9750 M^3 \rb-9465 M^2 \rb^2+1295 M \rb^3+842 \rb^4)+3 a^2 M^2 \rb^{10} (2478 M^4 \\
&& \quad
 -\:5529 M^3 \rb +625 M^2 \rb^2+595 M \rb^3-21 \rb^4)+6 a M^{7/2} \rb^{23/2} (474
   M^2+433 M \rb-253 \rb^2) \\
&& \quad
 +\:3 M^3 \rb^{13} (-373 M^2+152 M \rb+\rb^2)
   \Big].
\end{IEEEeqnarray*}

% ---------------------------- huu Regularisation  ---------------------------------

\subsubsection{\texorpdfstring{$huu$}{huu} regularization}
The quantity
\begin{equation}
H^{\reg} = \frac12 h^\reg_{ab} u^a u^b,
\end{equation}
was first proposed by Detweiler \cite{Detweiler:2005kq} as a tool for constructing
gauge invariant measurements from self-force calculations. It has, since then, been proven
invaluable in extracting gauge invariant results from gauge dependent self-force
calculations \cite{Detweiler:2008ft,Barack:2011ed}.

Much the same as with self-force calculations, the calculation of $H^{\reg}$
requires the subtraction of the appropriate singular piece, $H^{\sing} = \frac12
h^{\sing}_{ab} u^a u^b$, from the full retarded field. In this section, we give
this subtraction in the form of mode-sum regularization parameters. In doing so,
we keep with our convention that the term proportional to $\ell+\tfrac12$ is
denoted by $H_{\lnpow{1}}$ ($=0$ in this case), the constant term is denoted by
$H_{[0]}$, and so on.

Note that, as in the self-force case, an ambiguity arises here due to the
presence of terms involving the four-velocity at $x$. One is free to arbitrarily
choose how to define this, provided $\lim_{x\to\bar{x}} u^a = u^\ab$. As before,
we choose this in such a way that the Boyer-Lindquist components of the
four-velocity at $x$ are exactly those at $\xb$.  The regularization parameters are
then given by
\begin{equation}
H_{[0]} = \frac{2 \mathcal{K}}{\pi \left(\rb^2+L^2 + \frac{2 a^2 M}{\rb} + a^2 \right)^{1/2} },
\end{equation}
\begin{equation}
H_{[1]} = 0,
\end{equation}
\begin{equation}
H_{[2]} = \frac{ \left(H^{\mathcal{E}}_{[2]} \mathcal{E} + H^{\mathcal{K}}_{[2]} \mathcal{K} \right)}{3 \pi  \rb^7 \left(\rb^2+L^2 + \frac{2 a^2 M}{\rb} + a^2 \right)^{3/2} \left( 2 a^2 M + a^2 \rb + L^2 \rb \right)^3 } ,
\end{equation}
where
\begin{IEEEeqnarray*}{rCl}
H^{\mathcal{E}}_{[2]} &=& 
	\big(12 M \rb^5 a^{12}+92 M^2 \rb^4 a^{12}+264 M^3 \rb^3 a^{12}+336 M^4
   \rb^2 a^{12}+160 M^5 \rb a^{12} - 24 \rb^8 a^{10}-240 M \rb^7
   a^{10} \\
&& \quad
	-\: 1104 L^2 M^6 a^{10} - 1104 M^2 \rb^6 a^{10}-2880 M^3 \rb^5 a^{10} + 48 L^2
   M \rb^5 a^{10}-4272 M^4 \rb^4 a^{10}+230 L^2 M^2 \rb^4 a^{10} \\
&& \quad
	-\: 3264 M^5
   \rb^3 a^{10} + 96 L^2 M^3 \rb^3 a^{10}-960 M^6 \rb^2 a^{10}-1116 L^2 M^4
   \rb^2 a^{10}-2096 L^2 M^5 \rb a^{10} - 48 \rb^{10} a^8 \\
&& \quad
	-\: 420 M \rb^9
   a^8-120 L^2 \rb^8 a^8-1556 M^2 \rb^8 a^8-2872 M^3 \rb^7 a^8 - 882 L^2 M
   \rb^7 a^8-2448 M^4 \rb^6 a^8-2781 L^2 M^2 \rb^6 a^8 \\
&& \quad
	-\: 672 M^5 \rb^5
   a^8 - 4770 L^2 M^3 \rb^5 a^8+72 L^4 M \rb^5 a^8-4272 L^2 M^4 \rb^4 a^8+90
   L^4 M^2 \rb^4 a^8 - 1440 L^2 M^5 \rb^3 a^8 \\
&& \quad
	-\: 1044 L^4 M^3 \rb^3 a^8-2928
   L^4 M^4 \rb^2 a^8-2112 L^4 M^5 \rb a^8 - 24 \rb^{12} a^6-168 M
   \rb^{11} a^6-195 L^2 \rb^{10} a^6-456 M^2 \rb^{10} a^6 \\
&& \quad
	-\: 480 M^3
   \rb^9 a^6 - 1086 L^2 M \rb^9 a^6-240 L^4 \rb^8 a^6-96 M^4 \rb^8
   a^6-2119 L^2 M^2 \rb^8 a^6 - 1528 L^2 M^3 \rb^7 a^6 \\
&& \quad
	-\: 1098 L^4 M \rb^7
   a^6-84 L^2 M^4 \rb^6 a^6-1578 L^4 M^2 \rb^6 a^6 - 696 L^4 M^3 \rb^5
   a^6+48 L^6 M \rb^5 a^6+84 L^4 M^4 \rb^4 a^6 \\
&& \quad
	-\: 190 L^6 M^2 \rb^4 a^6 - 1320
   L^6 M^3 \rb^3 a^6-1476 L^6 M^4 \rb^2 a^6-75 L^2 \rb^{12} a^4-246 L^2 M
   \rb^{11} a^4 - 297 L^4 \rb^{10} a^4 \\
&& \quad
	-\: 84 L^2 M^2 \rb^{10} a^4+216 L^2 M^3
   \rb^9 a^4-690 L^4 M \rb^9 a^4-240 L^6 \rb^8 a^4 + 529 L^4 M^2 \rb^8
   a^4+1374 L^4 M^3 \rb^7 a^4 \\
&& \quad
	-\: 402 L^6 M \rb^7 a^4+771 L^6 M^2 \rb^6
   a^4 + 1194 L^6 M^3 \rb^5 a^4+12 L^8 M \rb^5 a^4-190 L^8 M^2 \rb^4 a^4-444
   L^8 M^3 \rb^3 a^4 \\
&& \quad
	-\:  78 L^4 \rb^{12} a^2+36 L^4 M \rb^{11} a^2-201 L^6
   \rb^{10} a^2+384 L^4 M^2 \rb^{10} a^2+198 L^6 M \rb^9 a^2 - 120 L^8
   \rb^8 a^2+1092 L^6 M^2 \rb^8 a^2 \\
&& \quad
	+\: 162 L^8 M \rb^7 a^2+672 L^8 M^2
   \rb^6 a^2 - 48 L^{10} M^2 \rb^4 a^2 -27 L^6 \rb^{12}+114 L^6 M \rb^{11}-51 L^8
   \rb^{10}+222 L^8 M \rb^9 \\
&& \quad
	-\: 24 L^{10} \rb^8+108 L^{10} M \rb^7 \big) \\
&&
	+\: 4 a E L M \big(-71 a^4
   \rb^{11}-75 L^4 \rb^{11}-146 a^2 L^2 \rb^{11}-262 a^4 M
   \rb^{10}-270 a^2 L^2 M \rb^{10} - 147 a^6 \rb^9-159 L^6 \rb^9 \\
&& \quad
	-\: 465
   a^2 L^4 \rb^9-453 a^4 L^2 \rb^9-240 a^4 M^2 \rb^9-811 a^6 M
   \rb^8 - 873 a^2 L^4 M \rb^8-1684 a^4 L^2 M \rb^8-76 a^8 \rb^7-78
   L^8 \rb^7 \\
&& \quad
	-\: 310 a^2 L^6 \rb^7 - 462 a^4 L^4 \rb^7-306 a^6 L^2
   \rb^7-1490 a^6 M^2 \rb^7-1542 a^4 L^2 M^2 \rb^7-912 a^6 M^3
   \rb^6 - 522 a^8 M \rb^6 \\
&& \quad
	-\: 549 a^2 L^6 M \rb^6-1620 a^4 L^4 M
   \rb^6-1593 a^6 L^2 M \rb^6 - 1323 a^8 M^2 \rb^5-1368 a^4 L^4 M^2
   \rb^5-2691 a^6 L^2 M^2 \rb^5 \\
&& \quad
	-\: 1460 a^8 M^3 \rb^4 - 1458 a^6 L^2 M^3
   \rb^4+37 a^{10} M \rb^4+24 a^2 L^8 M \rb^4+109 a^4 L^6 M
   \rb^4+183 a^6 L^4 M \rb^4 \\
&& \quad
	+\:  135 a^8 L^2 M \rb^4-588 a^8 M^4
   \rb^3+291 a^{10} M^2 \rb^3+222 a^4 L^6 M^2 \rb^3 + 735 a^6 L^4 M^2
   \rb^3+804 a^8 L^2 M^2 \rb^3 \\
&& \quad
	+\: 858 a^{10} M^3 \rb^2+738 a^6 L^4 M^3
   \rb^2 + 1596 a^8 L^2 M^3 \rb^2+1124 a^{10} M^4 \rb+1056 a^8 L^2 M^4
   \rb+552 a^{10} M^5\big) \\
&& 
	-\: 3 E^2 \big(368 M^6 a^{12}+4 M \rb^5 a^{12}+55 M^2 \rb^4 a^{12}+280 M^3
   \rb^3 a^{12}+680 M^4 \rb^2 a^{12} + 800 M^5 \rb a^{12}-9 \rb^8
   a^{10} \\
&& \quad
	-\: 124 M \rb^7 a^{10}-604 M^2 \rb^6 a^{10}-1368 M^3 \rb^5 a^{10} + 12
   L^2 M \rb^5 a^{10}-1472 M^4 \rb^4 a^{10}+160 L^2 M^2 \rb^4 a^{10} \\
&& \quad
	-\: 608
   M^5 \rb^3 a^{10} + 672 L^2 M^3 \rb^3 a^{10}+1152 L^2 M^4 \rb^2 a^{10}+704
   L^2 M^5 \rb a^{10}-18 \rb^{10} a^8 - 224 M \rb^9 a^8-35 L^2 \rb^8
   a^8 \\
&& \quad	
	-\: 901 M^2 \rb^8 a^8-1492 M^3 \rb^7 a^8-438 L^2 M \rb^7 a^8 - 884 M^4
   \rb^6 a^8-1685 L^2 M^2 \rb^6 a^8-2604 L^2 M^3 \rb^5 a^8 \\
&& \quad
	+\: 12 L^4 M
   \rb^5 a^8 -  1412 L^2 M^4 \rb^4 a^8+171 L^4 M^2 \rb^4 a^8+540 L^4 M^3
   \rb^3 a^8+492 L^4 M^4 \rb^2 a^8 - 9 \rb^{12} a^6-96 M \rb^{11}
   a^6 \\
&& \quad
	-\: 55 L^2 \rb^{10} a^6-278 M^2 \rb^{10} a^6-244 M^3 \rb^9 a^6 - 596 L^2
   M \rb^9 a^6-51 L^4 \rb^8 a^6-1679 L^2 M^2 \rb^8 a^6-1414 L^2 M^3
   \rb^7 a^6 \\
&& \quad
	-\: 572 L^4 M \rb^7 a^6-1567 L^4 M^2 \rb^6 a^6-1254 L^4 M^3
   \rb^5 a^6+4 L^6 M \rb^5 a^6 + 82 L^6 M^2 \rb^4 a^6+148 L^6 M^3
   \rb^3 a^6 \\
&& \quad
	-\: 20 L^2 \rb^{12} a^4-176 L^2 M \rb^{11} a^4-56 L^4
   \rb^{10} a^4 - 266 L^2 M^2 \rb^{10} a^4-512 L^4 M \rb^9 a^4-33 L^6
   \rb^8 a^4-774 L^4 M^2 \rb^8 a^4 \\
&& \quad
	-\:  326 L^6 M \rb^7 a^4-486 L^6 M^2
   \rb^6 a^4+16 L^8 M^2 \rb^4 a^4-13 L^4 \rb^{12} a^2-80 L^4 M
   \rb^{11} a^2 - 19 L^6 \rb^{10} a^2-140 L^6 M \rb^9 a^2 \\
&& \quad
	-\: 8 L^8 \rb^8
   a^2-68 L^8 M \rb^7 a^2-2 L^6 \rb^{12}\big),
\end{IEEEeqnarray*}
\begin{IEEEeqnarray*}{rCl}
H^{\mathcal{K}}_{[2]} &=& 
	\rb^3 \big(24 \rb^5 a^{10}+180 M \rb^4 a^{10}+512
   M^2 \rb^3 a^{10}+656 M^3 \rb^2 a^{10}+320 M^4 \rb a^{10} + 93 \rb^7
   a^8+624 M \rb^6 a^8-912 L^2 M^5 a^8 \\
&& \quad
	+\: 120 L^2 \rb^5 a^8+1448 M^2 \rb^5
   a^8 + 976 M^3 \rb^4 a^8+636 L^2 M \rb^4 a^8-912 M^4 \rb^3 a^8+891 L^2 M^2
   \rb^3 a^8-1152 M^5 \rb^2 a^8 \\
&& \quad
	-\: 434 L^2 M^3 \rb^2 a^8-1720 L^2 M^4
   \rb a^8+69 \rb^9 a^6+354 M \rb^8 a^6+375 L^2 \rb^7 a^6 + 492 M^2
   \rb^7 a^6-168 M^3 \rb^6 a^6 \\
&& \quad
	+\: 1620 L^2 M \rb^6 a^6+240 L^4 \rb^5
   a^6-576 M^4 \rb^5 a^6 + 1207 L^2 M^2 \rb^5 a^6-2720 L^2 M^3 \rb^4 a^6+732
   L^4 M \rb^4 a^6 \\
&& \quad
	-\: 3372 L^2 M^4 \rb^3 a^6 - 450 L^4 M^2 \rb^3 a^6-2896 L^4
   M^3 \rb^2 a^6-2052 L^4 M^4 \rb a^6+210 L^2 \rb^9 a^4 + 534 L^2 M
   \rb^8 a^4 \\
&& \quad
	+\: 567 L^4 \rb^7 a^4-384 L^2 M^2 \rb^7 a^4-1224 L^2 M^3
   \rb^6 a^4 +  1074 L^4 M \rb^6 a^4+240 L^6 \rb^5 a^4-2017 L^4 M^2
   \rb^5 a^4 \\
&& \quad
	-\: 3726 L^4 M^3 \rb^4 a^4 + 180 L^6 M \rb^4 a^4-1525 L^6 M^2
   \rb^3 a^4-1806 L^6 M^3 \rb^2 a^4+213 L^4 \rb^9 a^2 - 18 L^4 M \rb^8
   a^2 \\
&& \quad
	+\: 381 L^6 \rb^7 a^2-888 L^4 M^2 \rb^7 a^2-216 L^6 M \rb^6 a^2+120 L^8
   \rb^5 a^2 - 1776 L^6 M^2 \rb^5 a^2-192 L^8 M \rb^4 a^2 \\
&& \quad
	-\: 696 L^8 M^2
   \rb^3 a^2+72 L^6 \rb^9-198 L^6 M \rb^8 + 96 L^8 \rb^7-294 L^8 M
   \rb^6+24 L^{10} \rb^5-96 L^{10} M \rb^4\big) \\
&&
	+\: 4 a  \rb^3 E L M \big(456 M^4 a^8+45 \rb^4 a^8+322 M \rb^3 a^8+862 M^2
   \rb^2 a^8+1024 M^3 \rb a^8 + 116 \rb^6 a^6+636 M \rb^5 a^6 \\
&& \quad
	+\: 183 L^2
   \rb^4 a^6+1162 M^2 \rb^4 a^6+708 M^3 \rb^3 a^6 + 992 L^2 M \rb^3
   a^6+1765 L^2 M^2 \rb^2 a^6+1026 L^2 M^3 \rb a^6+71 \rb^8 a^4 \\
&& \quad
	+\: 262 M
   \rb^7 a^4 + 358 L^2 \rb^6 a^4+240 M^2 \rb^6 a^4+1326 L^2 M \rb^5
   a^4+279 L^4 \rb^4 a^4+1206 L^2 M^2 \rb^4 a^4 + 1018 L^4 M \rb^3 a^4 \\
&& \quad
	+\: 903
   L^4 M^2 \rb^2 a^4+146 L^2 \rb^8 a^2+270 L^2 M \rb^7 a^2+368 L^4
   \rb^6 a^2 + 690 L^4 M \rb^5 a^2+189 L^6 \rb^4 a^2+348 L^6 M \rb^3
   a^2 \\
&& \quad
	+\: 75 L^4 \rb^8+126 L^6 \rb^6+48 L^8 \rb^4\big) \\
&&
	-\: 3 \rb^3 E^2 \big(304 M^5 a^{10}+8 \rb^5 a^{10}+90 M \rb^4 a^{10}+386 M^2
   \rb^3 a^{10}+796 M^3 \rb^2 a^{10} + 792 M^4 \rb a^{10}+32 \rb^7
   a^8 \\
&& \quad
	+\: 308 M \rb^6 a^8+32 L^2 \rb^5 a^8+1068 M^2 \rb^5 a^8 + 1600 M^3
   \rb^4 a^8+304 L^2 M \rb^4 a^8+880 M^4 \rb^3 a^8+1003 L^2 M^2
   \rb^3 a^8 \\
&& \quad
	+\:  1388 L^2 M^3 \rb^2 a^8+684 L^2 M^4 \rb a^8+24 \rb^9
   a^6+186 M \rb^8 a^6+113 L^2 \rb^7 a^6 + 458 M^2 \rb^7 a^6+364 M^3
   \rb^6 a^6 \\
&& \quad
	+\: 854 L^2 M \rb^6 a^6+48 L^4 \rb^5 a^6+2017 L^2 M^2 \rb^5
   a^6 + 1522 L^2 M^3 \rb^4 a^6+370 L^4 M \rb^4 a^6+849 L^4 M^2 \rb^3
   a^6 \\
&& \quad
	+\: 602 L^4 M^3 \rb^2 a^6 + 65 L^2 \rb^9 a^4+356 L^2 M \rb^8 a^4+146 L^4
   \rb^7 a^4+446 L^2 M^2 \rb^7 a^4+776 L^4 M \rb^6 a^4 + 32 L^6 \rb^5
   a^4 \\
&& \quad
	+\: 945 L^4 M^2 \rb^5 a^4+188 L^6 M \rb^4 a^4+232 L^6 M^2 \rb^3 a^4+58
   L^4 \rb^9 a^2 + 170 L^4 M \rb^8 a^2+81 L^6 \rb^7 a^2+230 L^6 M
   \rb^6 a^2 \\
&& \quad
	+\: 8 L^8 \rb^5 a^2+32 L^8 M \rb^4 a^2 + 17 L^6 \rb^9+16 L^8
   \rb^7\big) .
\end{IEEEeqnarray*}

\subsection{Example}
\begin{figure}
\includegraphics[width=10cm]{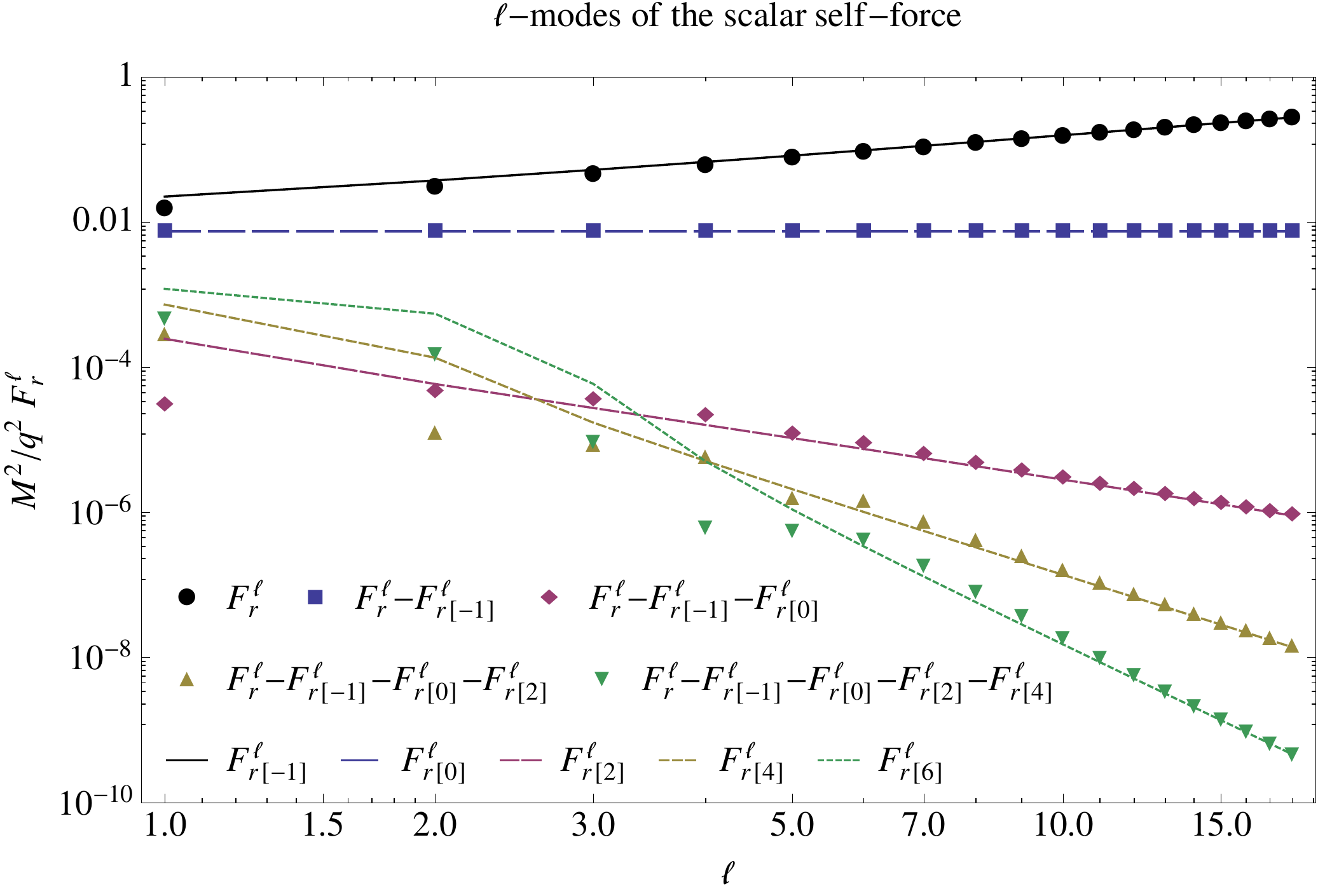}
\caption{Regularization of the $\ell$-modes of the radial component of the scalar
self-force in Kerr spacetime for the case of a particle in an eccentric,
equatorial geodesic orbit with $E = 0.955492$ and $L = 3.59656M$. The background
Kerr black hole has spin $a = 0.5M$. In decreasing slope the lines represent the
regularization parameters, $F^{\ell}_{r[-1]}$, $F^{\ell}_{r[0]}$, $F^{\ell}_{r[2]}$,
$F^{\ell}_{r[4]}$ and $F^{\ell}_{r[6]}$ (this last parameter is computed using a numerical fit to
the modes $\ell = 15, \ldots, 18$). The dots give the self-force computed from a frequency domain
calculation of the retarded field and regularized by
subtracting in turn the cumulative sum of $F^{\ell}_{r[-1]}$,
$F^{\ell}_{r[0]}$, $F^{\ell}_{r[2]}$ and $F^{\ell}_{r[4]}$.}
\label{fig:l-modes}
\end{figure}

To illustrate the effectiveness of the higher-order regularization parameters,
we consider, as an example, the case of a scalar charge on an eccentric geodesic
orbit, with $E=0.955492$ and $L=3.59656M$, in a Kerr spacetime with $a=0.5M$. The
self-force, in this case, was computed in Ref.~\cite{Warburton:2011hp} using
a frequency domain calculation of the retarded field in combination with the
first two regularization parameters. Figure \ref{fig:l-modes} shows the effect
of using higher order regularization parameters in this calculation. As
expected, the numerical $\ell$-modes computed in Ref.~\cite{Warburton:2011hp}
asymptotically fall off as $\ell^{-2}$ after subtracting the leading two parameters.
Our $F^{\ell}_{r[2]}$ regularization parameter
analytically gives the coefficient of this subleading order in $1/\ell$ behavior.
After subtracting this leading order behavior from the numerical modes, we find
a remainder that falls off as $\ell^{-4}$, as expected, with the coefficient given
analytically by our $F^{\ell}_{r[4]}$ regularization parameter. Upon further subtraction of
$F^{\ell}_{r[4]}$, the remainder falls off as $\ell^{-6}$, as anticipated.

% -------------------------------------------------------------------------------
% -------------------- Effective Source and m-mode Regularization ---------------------
% -------------------------------------------------------------------------------

\section{Effective Source and \texorpdfstring{$m$}{m}-mode Regularization}
\label{sec:m-mode}

% -------------------------------------------------------------------------------
% ------------------------------ Effective Source  ----------------------------------
% -------------------------------------------------------------------------------
\subsection{Effective source approach to the self-force}

The effective source approach -- independently proposed by Barack and Golbourn
\cite{Barack:2007jh} and by Vega and Detweiler \cite{Vega:2007mc}
-- relies on knowledge of the singular field to derive an equation for a
regularized field that gives the self-force without any need for postprocessed
regularization. If the singular field is known exactly, then the regularized
field is totally regular and is a solution of the homogeneous wave equation. In
reality, exact expressions for the singular field can be obtained only for very
simple spacetimes. More generally, the best one can do is an approximation such
as the one derived in this paper. Splitting the retarded field into approximate
singular and regularized parts (where a tilde denotes an approximation valid in the neighborhood of the world line),
\begin{equation}
\varphi^A_{\rm \ret} = \tilde{\varphi}^A_{\rm \sing} + \tilde{\varphi}^A_{\rm \reg},
\end{equation}
substituting into the wave equation, Eq.~(2.1) of Paper I, and rearranging, we
obtain an equation for the regularized field,
\begin{equation}
\mathcal{D}^{A}{}_B \tilde{\varphi}^{B}_{\rm \reg} = S_{\rm eff}^A,
\end{equation}
with an \emph{effective source},
\begin{equation}
S_{\rm eff}^{A} = -\mathcal{D}^{A}{}_B \tilde{\varphi}^{B}_{\rm \sing} - 4\pi \mathcal{Q} \int u^{A} \delta_4 \left( x,z(\tau') \right) d\tau'.
\end{equation}
For sufficiently good approximations to the singular field,
$\tilde{\varphi}^A_{\rm \reg}$ and $S^A$ are finite everywhere, in particular, on
the world line. As a result, one never encounters problematic singularities or
$\delta$ functions, making the approach particularly suitable for use in time
domain numerical simulations. A detailed review of this approach can be found in
\cite{Vega:2011wf,Wardell:2011gb}.

One disadvantage of the effective source approach stems from the fact that the
source must be evaluated in an extended region around the world line. Since the
source is derived from a complicated expansion approximating the singular field,
its evaluation can dominate the run time of a numerical code. This problem is
exasperated as increasingly good approximations to the singular field -- using
increasingly high-order series expansions -- are used, placing a practical
upper limit on the order of the singular field approximation that may be used in
effective source calculations. Existing calculations
\cite{Dolan:2010mt,Dolan:2011dx,Diener:2011cc} settled
on what appears to be a sweet spot, using an approximation accurate to
$\mathcal{O}(\epsilon^2)$.

Despite it being possible to compute higher order effective sources from our
singular field approximation, this practical consideration
may appear to rule out the usefulness of high-order expansions of the
singular field in effective source calculations. This is particularly so in the case of the
Kerr spacetime, where even an order $\mathcal{O}(\epsilon^2)$ approximation to
the singular field is quite unwieldy. However, it turns out that high-order
expansions can, in fact, be put to good use in effective source calculations. In
this section, we show how this may be achieved in the case of the $m$-mode
approach to effective source calculations. In this approach, one first performs
a decomposition into $m$-modes,
\begin{equation}
\label{eq:m-decomposition}
\tilde{\varphi}^A_{\rm \reg}{}^{(m)} = \frac{1}{2\pi}\int_{-\pi}^{\pi} \tilde{\varphi}^A_{\rm \reg} e^{-i m \phi} d\phi,
\end{equation}
and independently evolves the $m$-decomposed form of the wave equation for each
$m$-mode. These equations have an $m$ dependent effective source, which is
derived from the particular choice of approximation to the singular field. The
full field is then given as a sum of these individual modes,
\begin{equation}
\tilde{\varphi}^A_{\rm \reg} = \sum\limits_{m=-\infty}^{\infty} \tilde{\varphi}^A_{\rm \reg} {}^{(m)} e^{i m \phi_0},
\end{equation}
or equivalently,
\begin{equation}
\tilde{\varphi}^A_{\rm \reg} = \tilde{\varphi}^A_{\rm \reg}{}^{(0)} + 2 \sum\limits_{m=1}^{\infty} \mathrm{Re}[\tilde{\varphi}^A_{\rm \reg}{}^{(m)} e^{i m \phi_0}].
\end{equation}
For the remainder of this section, we will always work with these ``folded'' $m$-modes
and can therefore assume $m\ge0$.

For an approximation accurate to $\mathcal{O}(\epsilon^n)$, the numerical
solutions for the field fall off as $m^{-(n+2)}$ for $m$ even and as
$m^{-(n+3)}$ for $m$ odd. Obviously, only finitely many $m$-modes (typically
$\sim10$-$20$) can ever be computed numerically; with the error from truncating
the sum at a finite $m$ putting an upper limit on the accuracy of the self-force
that can be computed. This may be mitigated, somewhat, by fitting for a large-$m$
tail, but that fit itself requires more modes and is only ever approximate.
Here, we propose a much better solution; that is to use the higher order terms in the
singular field (those that have not been used in computing the effective
source) to analytically derive expressions for the tail. In many ways, this is
analogous to the $\ell$-mode regularization scheme, where there is a large-$\ell$ tail
and one can compute $\ell$-mode regularization parameters.

% -------------------------------------------------------------------------------
% ------------------ Derivation of m-mode Regularization Parameters -------------------
% -------------------------------------------------------------------------------

\subsection{Derivation of \texorpdfstring{$m$}{m}-mode regularization parameters}
For clarity, we carry out the following derivation for a scalar field; however, extending this to the cases of higher spin is straightforward.  To derive analytic expressions for the large-$m$ tail, we first note that an
approximation to the singular field accurate to $\mathcal{O}(\epsilon^n)$ can be
written in the form
\begin{align}
\label{eq:full-phis-m-mode}
\Phi^{\rm S} (x)
  &= \frac{1}{\rho_0^{2n+3}} \Big[ \sum\limits_{\genfrac{}{}{0pt}{}{i=0}{i~\text{even}}}^{3(n+1)} A_{ni} \sin^i(\Delta\phi/2) + \sum\limits_{\genfrac{}{}{0pt}{}{i=0}{i~\text{odd}}}^{3(n+1)} A_{ni} \sin^{i-1}(\Delta\phi/2) \sin(\Delta\phi) \Big] + \mathcal{O}(\epsilon^{n+1})
\nonumber \\
  &= \frac{1}{\rho_0^{2n+3}} \Big[ \sum\limits_{\genfrac{}{}{0pt}{}{i=0}{i~\text{even}}}^{3(n+1)} A_{ni} \sin^i(\Delta\phi/2) + 2 \sum\limits_{\genfrac{}{}{0pt}{}{i=0}{i~\text{odd}}}^{3(n+1)} A_{ni} \sin^{i}(\Delta\phi/2) \cos(\Delta\phi/2) \Big] + \mathcal{O}(\epsilon^{n+1}),
\end{align}
where the coefficients $A_{ni}$ are functions of the world-line position, $r_0$ and $\theta_0$,
the constants of motion, $E$, $L$ and $C$, and $\Delta r$ and $\Delta
\theta$. This form has the benefit of ensuring that the approximation is regular
everywhere except on the world line, while still being amenable to analytic
integration in the $\phi$ direction. This makes it particularly appropriate for
use in $m$-mode effective source calculations \cite{Thornburg:Wardell}.

Using the leading orders [say, to $\mathcal{O}(\epsilon^p)$] in this expansion
to compute an effective source, one is left with a singular field remainder
that is finite, but of limited differentiability on the world line. Since it is
finite, we can safely set $\Delta r = \Delta \theta = 0$ in
Eq.~\eqref{eq:full-phis-m-mode}, leading to a singular field remainder that has
the form
\begin{align}
\label{eq:remainder-phis-m-mode}
\Phi^{\rm S} (x)
  &= \Big[2\Theta(\Delta\phi)-1\Big] \Big[ \sum\limits_{\genfrac{}{}{0pt}{}{i=p+1}{i~\text{odd}}}^{n} B_{ni} \sin^i(\Delta\phi/2) + 2 \sum\limits_{\genfrac{}{}{0pt}{}{i=p+1}{i~\text{even}}}^{n} B_{ni} \sin^{i}(\Delta\phi/2) \cos(\Delta\phi/2) \Big] + \mathcal{O}(\epsilon^{n+1}),
\end{align}
where $\Theta(\Delta\phi)$ is the Heaviside step function. Substituting this
into Eq.~\eqref{eq:m-decomposition} and noting that for even $j$
\begin{equation}
\int_{-\pi}^{\pi} \Big[2\Theta(\Delta\phi)-1\Big] \sin^j(\Delta\phi/2) \cos(\Delta\phi/2)  e^{-i m \phi} d\phi =
\frac{2 i m}{j+1} \int_{-\pi}^{\pi} \Big[2\Theta(\Delta\phi)-1\Big] \sin^{j+1}(\Delta\phi/2) e^{-i m \phi} d \phi,
\end{equation}
we are left with trivial integrals of the form
\begin{equation}
\int_{-\pi}^{\pi} \Big[2\Theta(\Delta\phi)-1\Big] \sin^{j+1}(\Delta\phi/2) e^{-i m \phi} d \phi
= \int_{-\pi}^{\pi} \Big[2\Theta(\Delta\phi)-1\Big] \sin^{j+1}(\Delta\phi/2) \cos(m \phi) d \phi.
\end{equation}
As a result, we see that the real-valued regularization parameters are given by the odd
terms in the expansion of the singular field and the imaginary-valued parameters are
given by the even terms. Furthermore, we see that the falloff with $m$ is
always an even power of $1/m$ in the real part and an odd power of $1/m$ in the
imaginary part.

While this analysis was done for the field, it should be noted that it equally
well applies to the self-force. The only modification necessary is to compute
the self-force from the singular field before setting $\Delta r = \Delta \theta
= 0$; the remainder of the calculation proceeds in exactly the same way.

Finally, we note that the $m$-mode regularization parameters, derived in this way, are
dependent on the singular field being written in the form given in
Eq.~\eqref{eq:full-phis-m-mode}. Effective source calculations may use some other
form for the approximation to the singular field (while still being accurate to
the same order), in which case, there is no guarantee that the regularization
parameters given here are appropriate.

% -------------------------------------------------------------------------------
% ---------------------- EM-mode Regularization Parameters -------------------------
% -------------------------------------------------------------------------------

\subsection{\texorpdfstring{$m$}{m}-mode regularization parameters}
Below, we give the results of applying this calculation to the scalar and gravitational cases. In doing so, we omit the
explicit dependence on $m$ that in each case is
\begin{gather}
F^m_{a[2]} = \frac{-4 F_{a[2]}}{\pi(2m-1)(2m+1)}, \quad
F^m_{a[4]} = \frac{24 F_{a[4]}}{\pi(2m-3)(2m-1)(2m+1)(2m+3)}, \nonumber \\
F^m_{a[6]} = \frac{-480 F_{a[6]}}{\pi(2m-5)(2m-3)(2m-1)(2m+1)(2m+3)(2m+5)}, \nonumber \\
F^m_{a[8]} = \frac{20160F_{a[8]}}{\pi(2m-7)(2m-5)(2m-3)(2m-1)(2m+1)(2m+3)(2m+5)(2m+7)}.
\end{gather}
As the expressions for generic orbits of Kerr spacetime are too large to be
of use in printed form, we give here only the representative expressions for two
cases: the $r$ component of the scalar self-force for a circular geodesic orbit
and the quantity $H = \tfrac12 h_{ab} u^a u^b$ in the gravitational, circular orbits case. We
direct the reader online \cite{barrywardell.net} for more generic expressions in
electronic form.

For circular orbits, the scalar self-force $m$-mode regularization parameters are:
\begin{IEEEeqnarray}{rCl}
F_{r [2]} &=& 
   \frac{M}{24 \rb^4 [a M+\rb \sqrt{M \rb}] [2 a \sqrt{M \rb}+\rb (\rb-3 M)]^{3/2} [a^2+\rb (\rb-2 M)]^{3/2}}
   \bigg[24 a^7 M^2
\nonumber \\ && \quad
    -\:24 a^6 M \sqrt{M \rb} (M-2 \rb) -4 a^5 M \rb (23 M^2+M \rb-6 \rb^2)+2 a^4 M \rb \sqrt{M \rb} (45 M^2-112 M \rb+31\rb^2)
\nonumber \\ && \quad
   +\:2 a^3 M \rb^2 (45 M^3+45 M^2 \rb-73 M \rb^2+19 \rb^3) -3 a^2 \rb^2 \sqrt{M \rb} (29 M^4-88 M^3 \rb+38 M^2 \rb^2-4 M \rb^3+\rb^4)
\nonumber \\ && \quad
    -\:6 a M \rb^4 (29 M^3-43 M^2 \rb+21 M \rb^2-3 \rb^3) -3 \rb^5 \sqrt{M \rb} (29 M^3-25 M^2\rb+3 M \rb^2+\rb^3) \bigg],
\end{IEEEeqnarray}
\begin{IEEEeqnarray}{rCl}
F_{r [4]} &=& 
   \frac{M^2}{1440 \rb^{9} [a M + \rb \sqrt{M \rb}][2 a \sqrt{M \rb}+\rb (\rb-3 M)]^{7/2} [a^2+\rb (\rb-2 M)]^{3/2} }
   \bigg[-23040 a^{14} M^2 \sqrt{M\rb}
\nonumber \\ && \quad
    +\:11520 a^{13} M^2 \rb (M-8 \rb)+384 a^{12} M \rb \sqrt{M \rb}(461 M^2-81 M \rb-360 \rb^2) -192 a^{11} M \rb^2(307 M^3
\nonumber \\ && \quad
    -\:3780 M^2 \rb+1233 M \rb^2+480 \rb^3)-64 a^{10} \rb^2 \sqrt{M \rb} (8549 M^4-3593 M^3 \rb-16212 M^2 \rb^2
\nonumber \\ && \quad
    +\:6336 M\rb^3+360 \rb^4) +32 a^9 M \rb^3 (2835 M^4-69401 M^3 \rb+46565 M^2 \rb^2+13779 M \rb^3-8748 \rb^4)
\nonumber \\ && \quad
    +\:192 a^8 \rb^3 \sqrt{M \rb}(4470 M^5-3621 M^4 \rb-15645 M^3 \rb^2+12662 M^2 \rb^3-1529 M \rb^4-342 \rb^5)
\nonumber \\ && \quad
    +\:16 a^7 \rb^4 (-1479 M^6+210966 M^5\rb-224760 M^4 \rb^2-49213 M^3 \rb^3+93619 M^2 \rb^4-19953 M \rb^5+180 \rb^6)
\nonumber \\ && \quad
    -\:16 a^6 \rb^4 \sqrt{M \rb} (43101 M^6-61443 M^5\rb-271980 M^4 \rb^2+343776 M^3 \rb^3-100489 M^2 \rb^4-7763 M \rb^5
\nonumber \\ && \quad
    +\:4158 \rb^6)+12 a^5 \rb^5 (-2367 M^7-221220 M^6\rb+337457 M^5 \rb^2+71894 M^4 \rb^3-262111 M^3 \rb^4+111498 M^2 \rb^5
\nonumber \\ && \quad
    -\:14459 M \rb^6+588 \rb^7) +3 a^4 \rb^5 \sqrt{M \rb}(76125 M^7-176307 M^6 \rb-1157559 M^5 \rb^2+1949709 M^4 \rb^3
\nonumber \\ && \quad
    -\:855873 M^3 \rb^4-26505 M^2 \rb^5+76235 M \rb^6-8065 \rb^7)+12 a^3\rb^7 (76125 M^7-152637 M^6 \rb-93174 M^5 \rb^2
\nonumber \\ && \quad
    +\:281414 M^4 \rb^3-166063 M^3 \rb^4+36555 M^2 \rb^5-3480 M \rb^6+460\rb^7) +18 a^2 \rb^8 \sqrt{M \rb} (76125 M^6
\nonumber \\ && \quad
    -\:145182 M^5 \rb+70771 M^4 \rb^2+16696 M^3 \rb^3-19905 M^2 \rb^4+3190 M \rb^5+225 \rb^6) +36 a \rb^{10} (25375 M^6
\nonumber \\ && \quad
    -\:47369 M^5 \rb+31856 M^4 \rb^2-8692 M^3 \rb^3+705 M^2 \rb^4-75 M \rb^5+40 \rb^6)
\nonumber \\ && \quad
    +\:9 \rb^{11} \sqrt{M\rb} (25375 M^5-47015 M^4 \rb+29014 M^3 \rb^2-4814 M^2 \rb^3-1365 M \rb^4+405 \rb^5)
   \bigg].
\end{IEEEeqnarray}
The gravitational, $m$-mode parameters for $H$ are
\begin{multline}
H_{[2]} =\\
 M^{3/2} \frac{44
   a^4 M+88 a^3 \sqrt{M \rb^3}-3 a^2 \rb (M-\rb) (29 M+15 \rb)+6 a \sqrt{M \rb^5} (14 \rb-29 M)-87 M \rb^4+45 \rb^5}{12
   \rb^{7/2} \left(a M+\sqrt{M\rb^3}\right) \left[a^2+\rb (\rb-2 M)\right]^{1/2}\left[2 a \sqrt{M \rb}+\rb (\rb-3 M)\right]^{1/2}},
\end{multline}
and
\begin{IEEEeqnarray}{rCl}
  H_{[4]} &=& \frac{M^{3/2}}{720 \rb^{15/2} (a
   M^{1/2}+\rb^{3/2}) [2 a (M \rb)^{1/2}+\rb (\rb-3 M)]^{7/2} [a^2+\rb (\rb-2 M)]^{1/2}} \times
\nonumber \\
&&
  \quad\Big[13824 a^{12} M^{5/2}+6912 a^{11} M^2 \sqrt{\rb} (3 M+8 \rb)-64 a^{10} M^{3/2} \rb (2249 M^2-2484 M \rb-1296 \rb^2)
\nonumber \\
&&
  \qquad -\:64 a^9 M \rb^{3/2} (1920 M^3+8383 M^2 \rb-6777 M \rb^2-864 \rb^3)
\nonumber \\
&&
  \qquad +\:48 a^8 M^{1/2} \rb^2 (11005 M^4-21094 M^3 \rb-12019 M^2 \rb^2+11664 M \rb^3+288 \rb^4)
\nonumber \\
&&
  \qquad +\:64 a^7 M \rb^{5/2} (3879 M^4+30408 M^3 \rb-44007 M^2 \rb^2+2981 M \rb^3+5562 \rb^4)
\nonumber \\
&&
  \qquad +\:4 a^6 M^{1/2} \rb^3 (-208989 M^5+544428 M^4 \rb+483978 M^3 \rb^2-880476 M^2 \rb^3+209395 M \rb^4+24192 \rb^5)
\nonumber \\
&&
  \qquad -\:12 a^5 \rb^{7/2} (14247 M^6+263427 M^5 \rb-515490 M^4 \rb^2+95446 M^3 \rb^3+154187 M^2 \rb^4-52041 M \rb^5-432 \rb^6)
\nonumber \\
&&
  \qquad +\:3 a^4 M^{1/2} \rb^4 (163125 M^6-528642 M^5 \rb-1218021 M^4 \rb^2+2583348 M^3 \rb^3-1176005 M^2 \rb^4+7790 M \rb^5
\nonumber \\
&&
  \qquad\quad+\:57445 \rb^6)
\nonumber \\
&&
  \qquad +\:12 a^3 \rb^{11/2} (163125 M^6-386172 M^5 \rb+19074 M^4 \rb^2+335148 M^3 \rb^3-201235 M^2 \rb^4+31920 M \rb^5
\nonumber \\
&&
  \qquad\quad +\:860 \rb^6)
\nonumber \\
&&
  \qquad +\:18 a^2 M^{1/2} \rb^7 (163125 M^5-337377 M^4 \rb+197294 M^3 \rb^2-2450 M^2 \rb^3-28315 M \rb^4+6075 \rb^5)
\nonumber \\
&&
  \qquad +\:36 a \rb^{17/2} (54375 M^5-103674 M^4 \rb+71932 M^3 \rb^2-20850 M^2 \rb^3+1725 M \rb^4+140 \rb^5)
\nonumber \\
&&
  \qquad +\:9 \sqrt{M} \rb^{10} (54375 M^4-97620 M^3 \rb+66074 M^2 \rb^2-20020 M \rb^3+2295 \rb^4)\Big].
\end{IEEEeqnarray}

\subsection{Example}
As an example application of these $m$-mode regularization parameters, we
consider the case of a scalar charge, on a circular geodesic orbit of radius
$10M$, in Kerr spacetime with $a=0.6M$. The self-force, in this case, was
computed in Ref.~\cite{Thornburg:Wardell}, using the $m$-mode effective source
approach, with an effective source derived from an approximation to the singular
field of the form \eqref{eq:full-phis-m-mode}, accurate to
$\mathcal{O}(\epsilon^2)$. As expected, this gives numerical results for the
$m$-modes of the self-force that asymptotically fall off as $m^{-4}$. In this
case, the $F^m_{r[2]}$ parameter is not needed as it has already been subtracted
through the effective source calculation. However, the $F^m_{r[4]}$ parameter
has not been subtracted and asymptotically gives the leading order behavior (in
$1/m$) of the modes. Subtracting this from the numerical results, therefore, leaves
a remainder that falls off as $m^{-6}$. Furthermore, a numerical fit of this
remainder can be done to numerically determine the next two parameters, in this
case, giving $F_{r[6]}=0.108797 q^2/M^2$ and $F_{r[8]}=11.3398 q^2/M^2$.

\begin{figure}
\includegraphics[width=10cm]{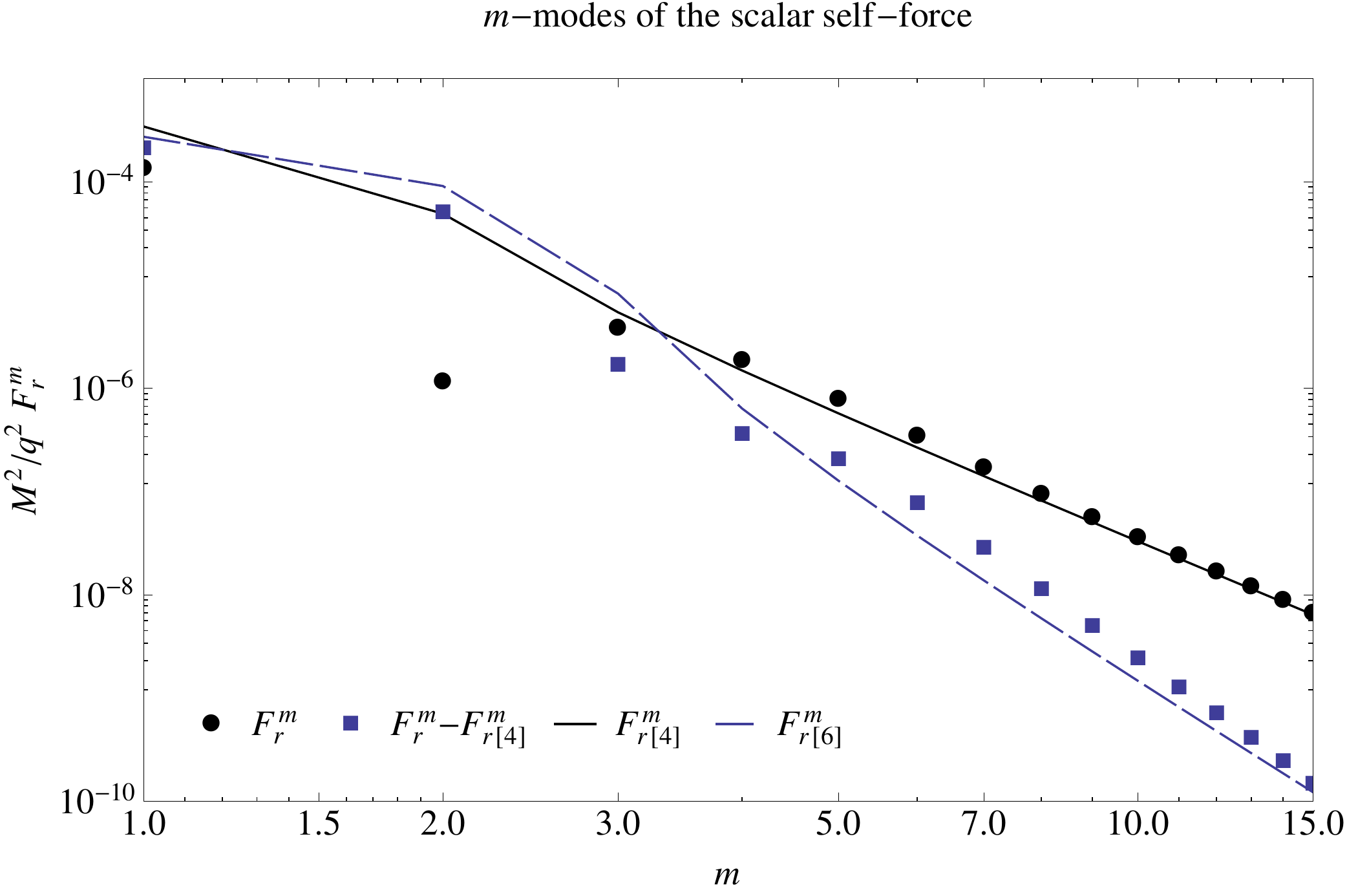}
\caption{Regularization of the radial component of the self-force for the case
of a scalar particle, on a circular geodesic of radius $r_0 = 10M$, in Kerr
spacetime with $a=0.6M$.  The numerical self-force modes
asymptotically match the $F^m_{r[4]}$ regularization parameter for
large $m$. After regularization, the remainder fall off as
$m^{-6}$, as expected.}
\label{fig:m-modes}
\end{figure}

In Fig.~\ref{fig:m-modes}, we plot the results of subtracting the analytic $F^m_{r[4]}$
and numerically fitted $F^m_{r[6]}$ regularization parameters, in turn, from the raw numerical data.
For large $m$, the numerical data falls off as $m^{-4}$, with the
coefficient matching our analytic prediction given by $F^m_{r[4]}$.
Subtracting this leading order behavior, we find that the remainder falls off
as $m^{-6}$, as expected.

\section{Discussion}
\label{sec:Discussion}
This paper extends the work of Paper I to the case of equatorial geodesic
orbits in Kerr spacetime. However, this only reflects a subset of the
possible geodesic orbits in that case. In general, geodesics of Kerr
spacetime do not lie in the equatorial plane. While our calculation could be
extended to cover the case of these more generic geodesics, we have chosen here
to restrict ourselves to the case of equatorial motion and work with the
significantly simpler expressions that ensue, leaving the more general case for
future work.

In our analysis, we have made use of scalar spherical harmonics that are not
particularly well suited to Kerr spacetime or gravitational perturbations.
A more appropriate choice of basis functions may be the spheroidal harmonics
for Kerr spacetime or the tensor harmonics in the gravitational case; it
may be more sensible to compute regularization parameters for these spheroidal
or tensor harmonic bases. However, from a practical perspective, most existing numerical
self-force calculations already make use of lower order versions of the
expressions given here. The example in Sec.~\ref{sec:ModeSum} clearly shows that these existing
calculations gain significant improvements in accuracy from the use of spherical
harmonic expansions. In this way, the end justifies the means: despite not being
a natural choice, the use of spherical harmonics is a good, practical choice.
Nevertheless, an adaptation of our calculation to the spheroidal or tensor harmonic basis,
and to other gauges, would make the results applicable in a much wider range of
contexts.

The Lorenz gauge metric perturbation equation on Kerr spacetime has not yet
been shown to be fully separable. It is likely that this would require the
development of tensor spheroidal harmonics, whose existence are, as-yet, unknown.
In the absence of these, it has not been possible to test the validity of our
electromagnetic and gravitational $\ell$-mode regularization parameters. However,
as in Paper I, deriving the expressions by independent methods gives
us strong confidence in our results. Another check is to set $a=0$, in which case,
the results agree with those of Paper I, which we know to be correct.

Note that the $m$-mode scheme is not affected by this issue as it is equally
applicable to both Schwarzschild and Kerr spacetimes, and is likewise equally as
valid in the gravitational, electromagnetic and scalar cases.  The only caveat
is that the $m$-mode regularization parameters are only guaranteed to be correct
for an effective source derived from a compatible approximation to the singular
field. Since there is a large amount of flexibility in the effective source
approach, if one chooses an incompatible singular field approximation, the
regularization parameters here must be modified appropriately.

\section*{Acknowledgements}
We are extremely grateful to Niels Warburton and Leor Barack for making
available their data for the unregularized $\ell$-modes of the retarded field, and
to Sam Dolan and Jonathan Thornburg for making their data available for the
$m$-modes of the retarded field. We also thank the participants of the 2011 and
2012 Capra meetings (in Southampton and the University of Maryland,
respectively) for many illuminating conversations.

A.H. has been supported by the Irish Research Council for Science, Engineering and
Technology, funded by the National Development Plan as well as the University College of Dublin's School of Mathematical Sciences. B.W. and A.C.O. gratefully
acknowledge support from Science Foundation Ireland under Grant No.
10/RFP/PHY2847.  Part of this work was supported by the COST Action MP0905
``Black Holes in a Violent Universe.''

\bibliography{references}{}

\end{document}